 \documentclass[structabstract]{aa}
 \usepackage{graphicx}
 \usepackage{txfonts}
\usepackage{longtable}
\usepackage{amssymb}
\usepackage{natbib}
\bibpunct{(}{)}{;}{a}{}{,}
\def\kms  {km~s$^{-1}$}

\def\wat {H$_2$O}
\def\meth {CH$_3$OH}
\def\Vlsr {\ifmmode {V_{\rm LSR}} \else {$V_{\rm LSR}$} \fi}
\def\Gp{G23.01$-$0.41}
\def\Gs{G16.59$-$0.05}
\def\IR{IRAS~20126$+$4104}
\def\AF{AFGL~5142}

\begin{document}


 \title{Unveiling the gas kinematics at 10~AU scales in high-mass star-forming regions} 
\subtitle{Milliarcsecond structure of 6.7~GHz methanol masers}

   \titlerunning{6.7~GHz maser internal gradients}

   \author{L. Moscadelli
          \inst{1}
          \and
          A. Sanna
         \inst{2}
          \and
          C. Goddi
          \inst{3}
          }

   \institute{INAF - Osservatorio Astrofisico di Arcetri,
              Largo E. Fermi 5, I-50125 Firenze, Italy\\
              \email{mosca@arcetri.astro.it}
             \and
                   Max-Planck-Institut f\"{u}r Radioastronomie,
                     Auf dem H\"{u}gel 69, 53121 Bonn, Germany\\
                \email{asanna@mpifr-bonn.mpg.de}
           \and
                     European Southern Observatory,
                     Karl-Schwarzschild-Strasse 2, D-85748 Garching bei M\"{u}nchen, Germany\\
              \email{cgoddi@eso.org}
             }



  \abstract
   { 
High-mass stars play a prominent role in Galactic evolution, 
but their formation mechanism is still poorly understood. 
This lack of knowledge reflects the observational limitations of present instruments, 
whose angular resolution (at the typical distances of massive protostars)
precludes probing circumstellar gas on scales of 1-100~AU, 
relevant for a detailed investigation of accretion structures 
and launch/collimation mechanims of outflows in high-mass star formation.
 }
   {This work presents a study of the milliarcsecond structure of the 6.7~GHz methanol masers at high-velocity resolution (0.09~\kms)
      in four high-mass star-forming regions: \Gs, \Gp, \IR, and \AF.}
   {We studied these sources
     by means of multi-epoch VLBI observations in the 22~GHz water and 6.7~GHz methanol masers, to determine the
     3-D~gas~kinematics within a few thousand AU from the (proto)star. Our results demonstrate the ability of maser emission
       to trace kinematic structures close to the (proto)star, revealing the presence of fast wide-angle and/or collimated outflows
       (traced by the \wat\ masers), and of rotation and infall (indicated by the \meth\ masers). The present work exploits the
       6.7~GHz maser data collected so far to investigate the milliarcsecond structure of this maser emission at high-velocity resolution.}
   {Most of the detected 6.7~GHz maser features present an ordered (linear, or arc-like) distribution
    of maser spots on the plane of the sky, together with a regular variation in the spot LSR velocity ($V_{\rm LSR}$)
   with position.
    Typical values for the amplitude of the \Vlsr gradients (defined in terms of the derivative of the spot \Vlsr with position)
      are found to be  \ 0.1--0.2~km~s$^{-1}$~mas$^{-1}$.
     In each of the four target sources, the orientation and the amplitude of most of the feature \Vlsr gradients
      remain remarkably stable in time, on timescales of (at least) several years.
      We also find that 
       the data are consistent with having the \Vlsr gradients and proper motion vectors
      in the same direction on the sky, considered the measurement uncertainties.
      In three (\Gs, \Gp, and \IR) of the four sources under examination,
       feature gradients with the best determined (sky-projected) orientation divide into two groups directed 
       approximately perpendicular to each other.}
     {The time persistency, the ordered   angular and spatial distribution,
      and the orientation generally similar to the proper motions,
      altogether suggest a kinematical interpretation for the origin
     of the 6.7~GHz maser \Vlsr gradients.
     This work shows that the organized motions (outflow, infall, and rotation) 
     revealed by the (22~GHz water and 6.7~GHz methanol) masers on large scales ($\sim$100--1000~AU)
     also persist to very small ($\sim$10~AU) scales.
     In this context, the present study demonstrates the potentiality of the mas-scale 6.7~GHz maser gradients  as a unique tool
     for investigating the gas kinematics on the smallest accessible scales in proximity to massive (proto)stars.
     }


   \maketitle

\section{Introduction}
\label{intr}


Intense maser transitions of the OH (at 1.6~and~6.0~GHz), \meth\ (at 6.7~and~12~GHz),
\wat\ (at 22~GHz), and SiO (at 43~GHz) molecules, are observed towards massive star-forming regions.
The elevate brightness temperature ($\ge$10$^7$~K) of the maser emissions permits us to observe
them with the Very Long Baseline Interferometry (VLBI) technique, achieving both
very high angular (1--10~mas) and velocity (0.1~\kms) resolution. Maser VLBI observations
are the unique mean by which one can explore the gas kinematics close (within tens or hundreds of AU)
to the forming high-mass (proto)star (e.g., \citealt{God05,San10a,San10b,Mat10,Mos11}),
and they have also been successfully used to infer the properties
of the magnetic field inside molecular cores \citep{Fis06b}.

VLBI maser observations are useful for testing the present theories of massive star formation.
To overcome the radiation pressure exerted by the already ignited star
and still enable mass accretion, most models point to the essential role that
accretion disks and jets should play. The former focus the ram pressure of the
accreting matter across the disk plane, and the latter channel stellar photons along the
jet axis and lower the radiation pressure across the equatorial plane. However, so far,
interferometric observations of {\it thermal} continuum and line emissions towards
massive star-forming regions have only identified a handful of disk candidates
in association with high-mass stars \citep[see e.g.,][]{Ces06}. In particular, no circumstellar
disks have been detected around early O-type stars, where only huge, massive, rotating structures
are seen, whose lifetimes appear to be less than the corresponding rotation periods \citep[see e.g.,][]{Bel11}.
One reason that could explain the difficulty detecting disks in massive star-forming regions
(with typical distances of a few kpc)
is that the angular resolution ($\ge1\arcsec$) achieved by present millimeter interferometers
could not suffice to resolve disks if  their size, as models predict, is comparable to or less than a few thousand AU.
On the contrary, supposing one can observe a masing transition from the disk gas,
the angular resolution achievable with maser VLBI observations is high enough to accurately
measure the rotation curve of the accretion disk.

For a decade we have been studying a sample of ten candidate high-mass (proto)stars
by means of VLBI observations of \wat\ 22~GHz, \meth\ 6.7~GHz, and OH 1.6~GHz masers
associated with the (proto)stellar environment. The \wat\ and \meth\ masers are observed at
several epochs to derive the maser proper motions, which, knowing the distance to the maser source,
can be translated into the sky-projected components of the maser velocity. Combining this information
with the line-of-sight velocity components, which are derived from the maser LSR velocity ($V_{\rm LSR}$)
 and the knowledge of the (proto)star systemic $V_{\rm LSR}$, a full 3-D picture of the motion of the masing gas can be obtained.
So far, data for four sources have been analyzed: \Gs\  \citep[hereafter SMC1]{San10a},
\Gp\ \citep[hereafter SMC2]{San10b}, \IR\ \citep[hereafter MCR]{Mos11}, and \AF\ \citep[hereafter GMS1 and GMS2, respectively]{God07,God11}.
These results demonstrate the ability of maser emission to trace kinematic structures close
to the (proto)star, revealing fast wide-angle and/or collimated outflows
(traced by the \wat\ masers) and rotation and infall (indicated by the \meth\ masers).

The present work exploits the 6.7~GHz maser data collected so far to investigate
the milliarcsecond structure of this maser emission at high-velocity resolution.
 Since the velocity resolution of the 6.7~GHz maser dataset
(0.09~km~s$^{-1}$) is high enough to resolve the maser emission linewidth
(with typical FWHM of \ 0.3~km~s$^{-1}$), by mapping each velocity channel
across the maser linewidth, one can investigate how the spatial structure of the emission
changes with the velocity. A similar study has been previously performed
for the intense 1.6~GHz \citep{Fis06a} and 6.0~GHz \citep{Fis07a} OH and 12~GHz
\meth\ \citep{Mos03} masers in the UC~H{\sc ii} region W3(OH). These observations have revealed
that it is quite common to observe a regular variation (along a line or an arc)
in the maser peak position with the \ $V_{\rm LSR}$.
In W3(OH), 1.6~and~6.0~GHz OH masers and 12~GHz \meth\ masers are found to have
similar amplitudes of the \Vlsr gradients
(calculated by dividing the maser \Vlsr linewidth by the path length
over which the maser peak position shifts on the plane of the sky),
varying in the range \ 0.01--1~km~s$^{-1}$~AU$^{-1}$.

This paper extends the study of the milliarcsecond structure of the maser emission
at high-velocity resolution to the 6.7~GHz \meth\ masers, presenting results in four distinct massive
star-forming regions: \Gs, \Gp, \IR, and \AF. Section~\ref{sum} summarizes
the main observational parameters of the EVN 6.7~GHz observations towards
these sources. Section~\ref{result} describes the basic characteristics of the
milliarcsecond, velocity structure of the 6.7~GHz masers, making use
of our multi-epoch observations to also examine the time variation.
Section~\ref{gra_pro} compares the directions on the sky plane of the maser \Vlsr gradients
and  proper motion vectors.
Section~\ref{dis} compares the results obtained among the four sources
and discusses a possible interpretation of the time-persistent velocity
gradients inside most of the 6.7~GHz maser cloudlets.


\section{Summary of \meth\ 6.7~GHz maser EVN observations}
\label{sum}

Using the European VLBI Network (EVN)\footnote{The European VLBI Network is a joint facility of European, Chinese,
South African, and other radio astronomy institutes founded by their national research councils.}, we
observed the sources \Gs, \Gp, \IR, and \AF\  in the 5$_{1}$-6$_{0}$~A$^{+}$
\meth\ transition (rest frequency 6.668519 GHz),
at three different epochs, over the years \ 2004--2009. Table~\ref{obs_evn} lists the EVN observing epochs
for each of the four sources, and gives the bibliographic reference of the articles where the results of these observations
have been published. We refer the reader to these articles both for a full description of the EVN observing setup
and for a discussion of the 3-D~gas~kinematics
close to the massive (proto)star as traced by the \wat\ 22~GHz and \meth\ 6.7~GHz masers.

To determine the 6.7~GHz maser absolute positions
and velocities, the EVN observations
were performed in phase-referencing mode, by fast switching between the maser source and one
strong, nearby calibrator. For each of the four maser targets,
the selected phase-reference calibrator (belonging to the
list of sources defining the International Celestial Reference Frame, ICRF)
has an angular separation on the sky from the maser less than \ $3\degr$.
In each of the four targets, the absolute position of the 6.7~GHz masers
is derived with an accuracy of a few milliarcseconds.

The antennae involved in the observations
were Medicina, Cambridge, Jodrell, Onsala, Effelsberg, Noto, Westerbork, Torun, Darnhall,
and Hartebeesthoek.
Because of technical problems, the Darnhall telescope replaced the Jodrell antenna
in the first epoch (November 2004) of the \IR\ and \AF\ EVN observations.
The Hartebeesthoek antenna took part in the observations only for the first two epochs
of  the low-DEC sources \Gs\ and \Gp. In the third epoch,
since the maser emission would be heavily resolved on the longest baselines involving the
Hartebeesthoek antenna, it was replaced
with the Onsala telescope.
All the EVN runs had a similar observing setup,
providing the same velocity resolution of \ 0.09~km~s$^{-1}$,
and a similar sensitivity, corresponding to a thermal rms noise level on the channel
maps of \ $\sim$10~mJy~beam$^{-1}$.

As indicated in Table~\ref{obs_evn}, the EVN observations
achieved comparable angular resolutions towards the four target sources,
with some larger beams for the two sources at low DEC, because of the
degraded $uv$-coverage. Isolated 6.7~GHz maser emission centers on a single-channel map
(named ``spots") are generally unresolved or slightly resolved with the EVN. By fitting
a two-dimensional elliptical Gaussian to the spot intensity distribution, the accuracy in the
relative positions of maser spots, $\delta\theta$,  can be calculated
through the formula \ $ \delta\theta = 0.5 \, \mathrm{FWHM} / \mathrm{SNR} $,
where FWHM is the fitted spot size (about the FWHM beam size for a compact source)
and SNR is the ratio of the spot intensity with the
image rms noise. Since most of the detected spots have intensities \ $\ge$1~Jy
and \ SNR$\ge$100, their relative positions are known with an accuracy
better than \ 0.1~mas.

\clearpage
\begin{table*}
\caption{\meth\ 6.7~GHz EVN observations}             
\label{obs_evn}      
\centering                          
\begin{tabular}{c c c c c c c c c}        
\hline\hline                 
\noalign{\smallskip}
Source & Epoch 1 & Epoch 2 & Epoch 3 &  & \multicolumn{2}{c}{Beam FWHM} & Vel. Res. & Ref. \\    
           &                  &                   &             & &  Maj. Axis & Min. Axis &        &     \\
           &   (yr m d)   &   (yr m d)  &  (yr m d)   & & (mas) & (mas) &  (\kms) &      \\
\noalign{\smallskip}
\hline                        
\noalign{\smallskip}
\Gs    & 2006 Feb 26   & 2007 Mar 16   & 2008 Mar 15     &    &  15  & 5 & 0.09  & SMC1   \\
\Gp   &  2006 Feb 27  & 2007  Mar 17   & 2008  Mar 16     &    &  13  & 5 & 0.09  & SMC2  \\
\IR     &  2004 Nov 06  & 2007 Mar 21  & 2009 Mar 11      &    &  8  & 6 & 0.09   & MCR      \\
\AF    & 2004 Nov 04 & 2007 Mar 16  & 2009 Mar 12       &     &  9  & 6 & 0.09   & GMS1, GMS2  \\
\noalign{\smallskip}
\hline   
\noalign{\smallskip}
\end{tabular}
\tablefoot{ \footnotesize Column~1 gives the maser source name, and Columns~2,~3,~and~4 
the observing date of the first, second, and third EVN epochs, respectively.
Columns~5~and~6 report the FWHM major and minor axes of the naturally-weighted beam
used to reconstruct the maser images. Column~7 indicates the velocity resolution of the
correlated visibilities, and Column~8 gives the bibliographic references.
}
\end{table*}

\section{Results}
\label{result}

In the following we use the term ``spot" to point to a compact emission center
on a single-channel map and the term ``feature" to refer to a collection of spots
emitting in contiguous channels at approximately the same position on the sky
(within the beam FWHM). In our view, a maser feature corresponds to a distinct,
masing cloud of gas, whose spatial and velocity structure is the subject of our analysis.

\subsection{6.7~GHz maser internal \Vlsr gradients}
\label{vgra}

\begin{figure*}

{\large \bf (a) }

\includegraphics[width=7.5cm]{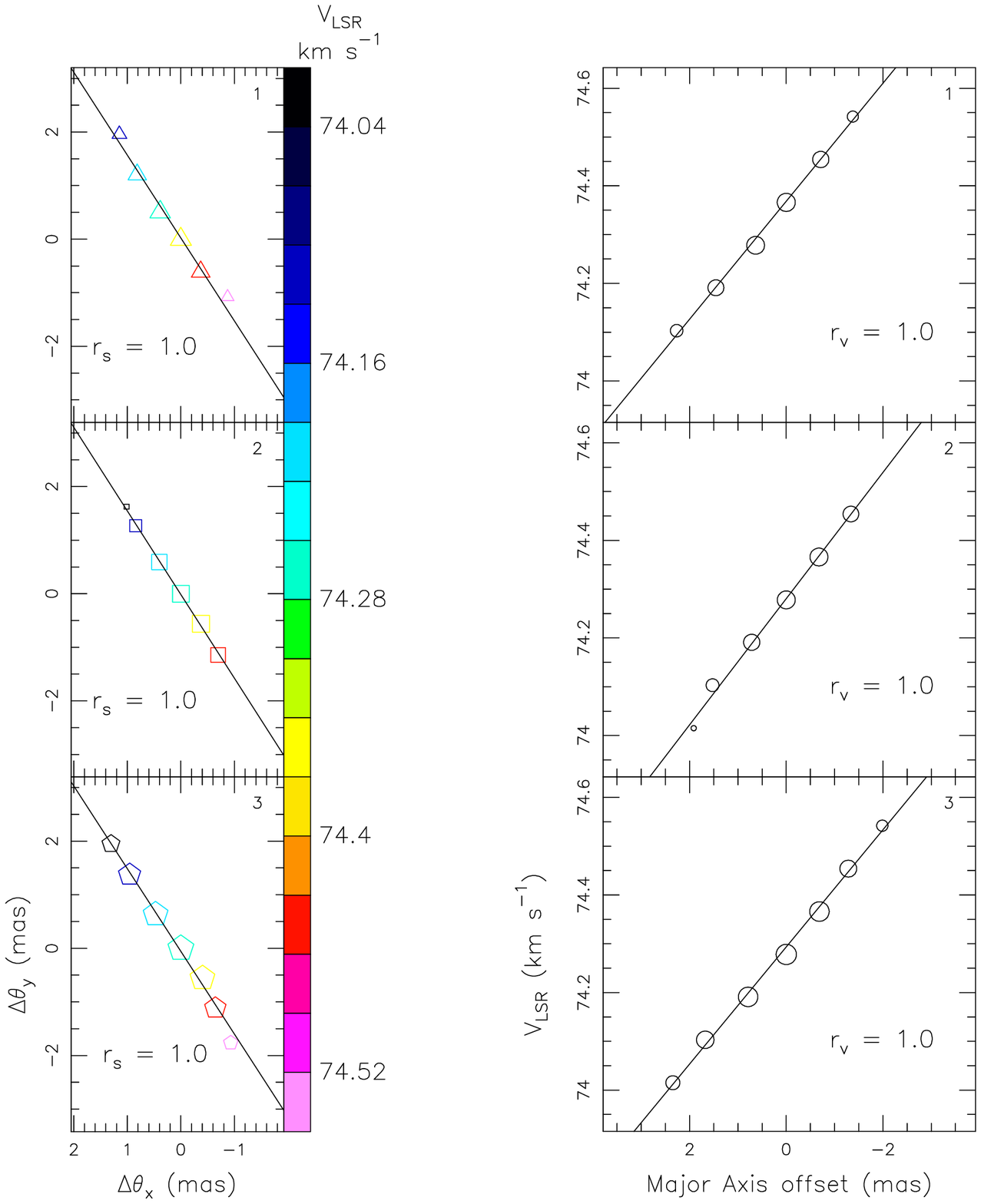}

{\large \bf (b) }

\includegraphics[width=7.5cm]{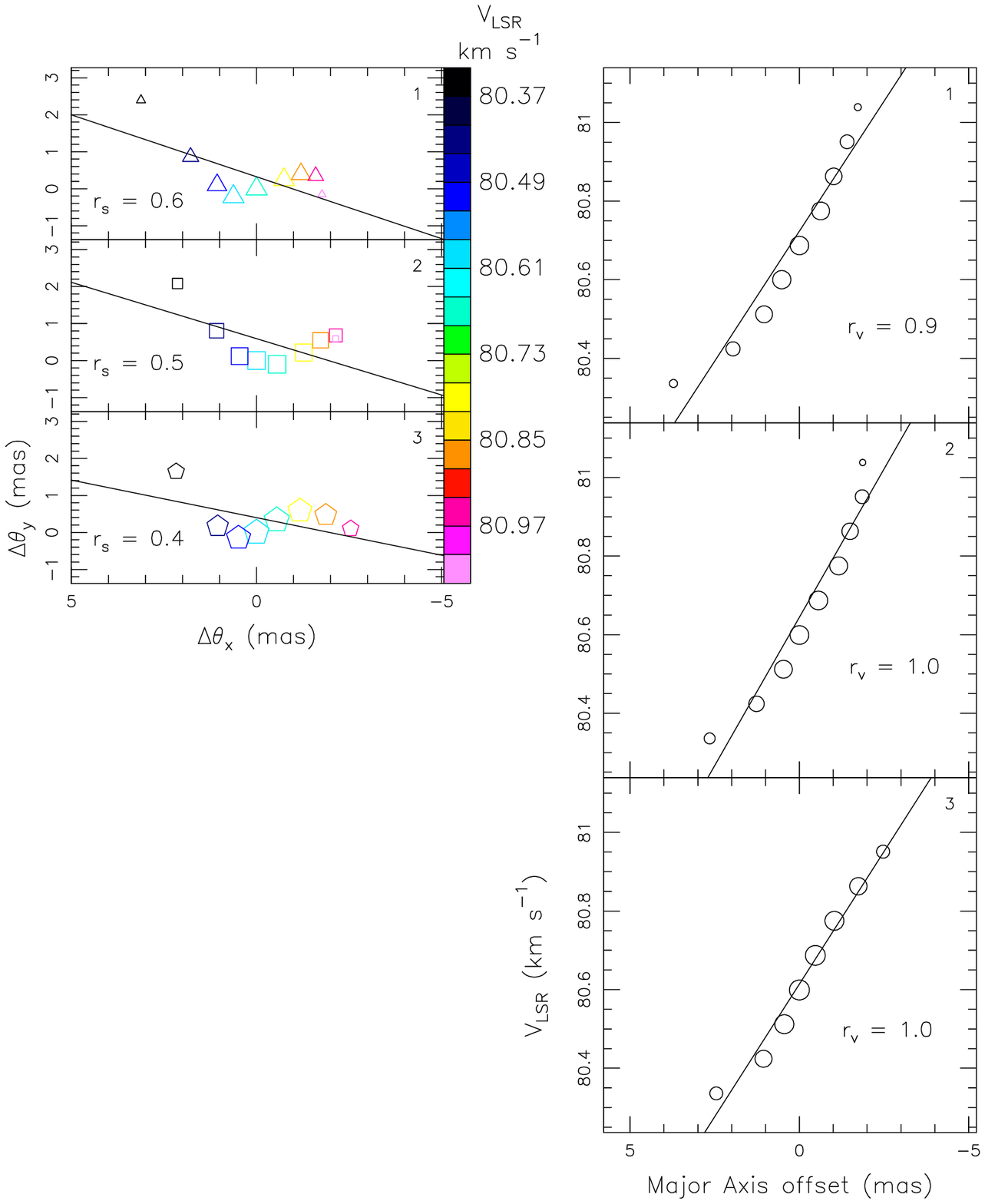}
\vspace{-0.2cm}
\caption{\tiny
Examples of  two 6.7~GHz maser features with ordered spatial and \Vlsr distribution.
{\large \bf(a)}~The upper, middle, and lower panels
refer to measurements at the first, second, and third observing epochs, respectively.
{\bf Left panels:} Spot spatial distribution.
{\it Triangles}, {\it squares}, and {\it pentagons} indicate the spot positions on the plane of the sky
at the first, second, and third epochs, respectively, with symbol size varying logarithmically
with the spot intensity. The symbol color indicates the spot $V_{\rm LSR}$, with the color-velocity
conversion code given in the wedge on the right side of the three panels. At each epoch,
the positions of the spots are relative to that of the epoch's most intense spot.
In each of the three panels,
the {\it continuous line} shows the (minimum $\chi^2$) linear  fit to the spot positions,
defining the direction of the major axis of the spot distribution.
The lower left corner of each panel reports the value of the linear correlation coefficient, $r_s$.
{\bf Right panels:} Spot $V_{\rm LSR}$--position distribution.
In each of the three panels, {\it empty circles} show the  plot of spot \Vlsr
versus the position offset along the major axis of the spot distribution.
Position offsets are taken as positive if increasing towards the east.
Circle size varies logarithmically with the spot intensity.
The
{\it continuous} lines show the (minimum $\chi^{2}$) linear  fit to the circle distribution.
The lower right corner of each panel reports the value
of the linear correlation coefficient, $r_v$.
{\large \bf(b)}~Same as for (a) for another feature of our sample.}
\label{feat_grad}
\end{figure*}

Figure~\ref{feat_grad} presents the spatial distribution and  the change in \Vlsr with position
of the spots belonging to two 6.7~GHz maser features, selected to be representative of the
ordered \Vlsr gradients commonly found inside the 6.7~GHz masers. 
The examples presented in Fig.~\ref{feat_grad} are two intense, persistent features from the source \Gp.
At each of the three observing epochs, spots from a given feature distribute in space close to a line and
show a regular variation in \Vlsr versus the position measured along the feature major axis (see also Fig.~6 in SMC1).
In the following discussion, we use \ $r_s$ \ and  \ $r_v$ \ to indicate the correlation coefficients of the linear fit to the
spot positions on the sky plane and to the \Vlsr variation with position
(measured along the major axis of the spot distribution), respectively.
For the maser feature in Fig.~\ref{feat_grad}a, the linear fits
of both the sky-projected spot distribution and the variation in  \Vlsr with positions
always have a correlation coefficient equal to 1.
For the feature in  Fig.~\ref{feat_grad}b, the spatial and
velocity distribution of the spots is less ordered, with correlation coefficients of
the linear fits in the range \  0.4--1.0.

The linear fits presented in Fig.~\ref{feat_grad} were performed for each feature
(of the four maser sources)
with three or more spots, thereby deriving the following quantities:  the P.A. of the feature
(major axis) orientation on the sky plane, $P_s$, with the degree of linear correlation
of the spot distribution on the sky plane measured by  \ $r_s$;  and the amplitude
of the  \Vlsr gradient with position, $\Gamma$ (in units of \ [km~s$^{-1}$~mas$^{-1}$]),
with the  degree of linear correlation of \Vlsr with position (measured along the
feature major axis) given by  \ $r_v$. With the adopted convention, that
position's offsets along the feature major axis are taken as positive if increasing to east,
\Vlsr increasing to east (west) results into positive (negative) velocity gradients.
Then, the P.A. of the feature orientation, $P_s$, varies in the range \ 0$\degr$--180$\degr$
for positive gradients, and in the range \ 180$\degr$--360$\degr$ \ for negative gradients.
The measurement errors of both the spot positions ($\le 0.1$~mas)
and \Vlsr ($\le 0.09$~km~s$^{-1}$) are small enough with respect to
the typical extension of the feature spot distribution (from a few mas to 10~mas)
and feature emission width
(0.5--1~km~s$^{-1}$) not to affect the precision with which the orientation
and the amplitude of the feature gradient can be determined.

\begin{figure*}
\includegraphics[width=7cm]{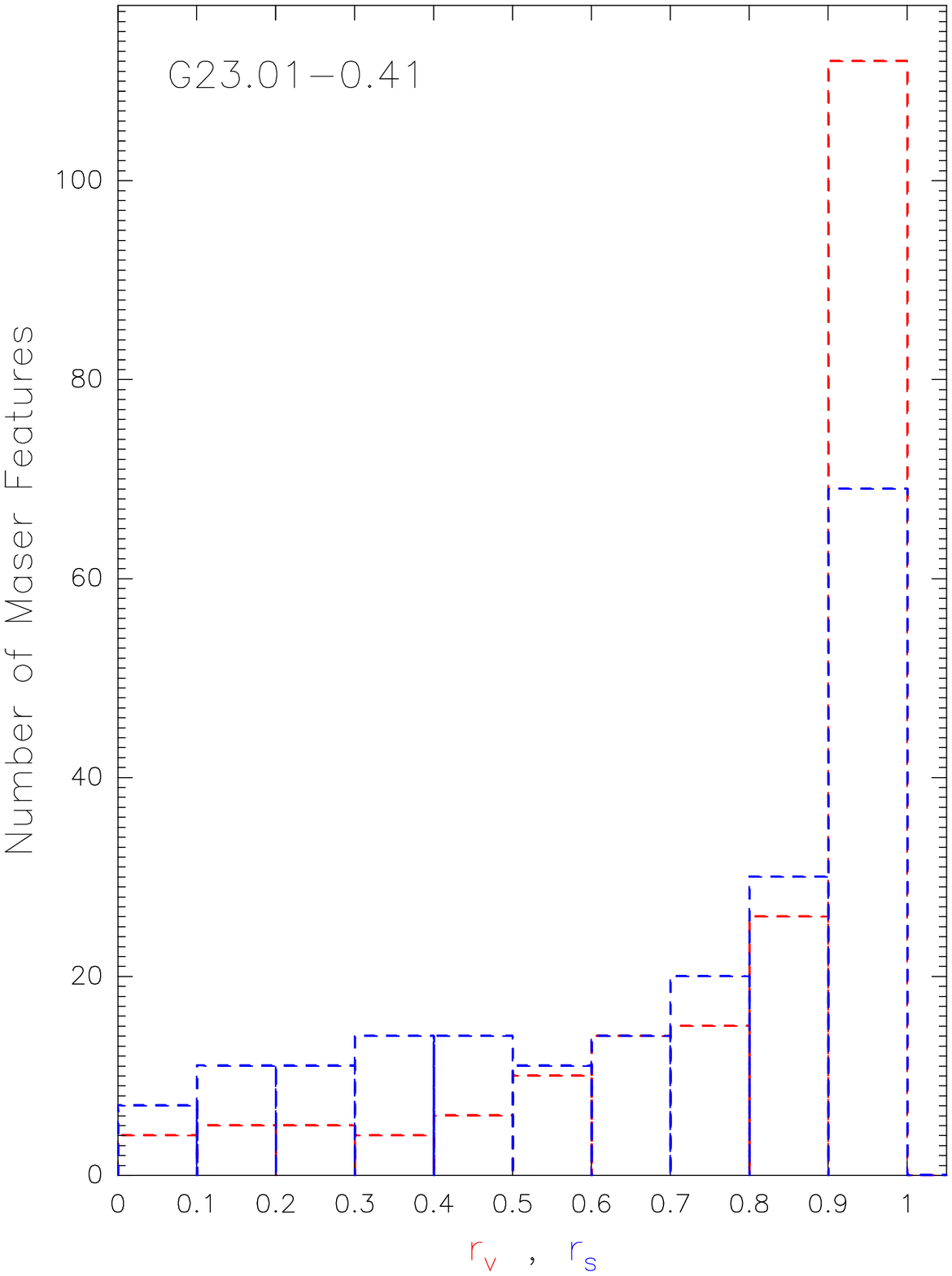}
\hspace{0.5cm}
\includegraphics[width=7cm]{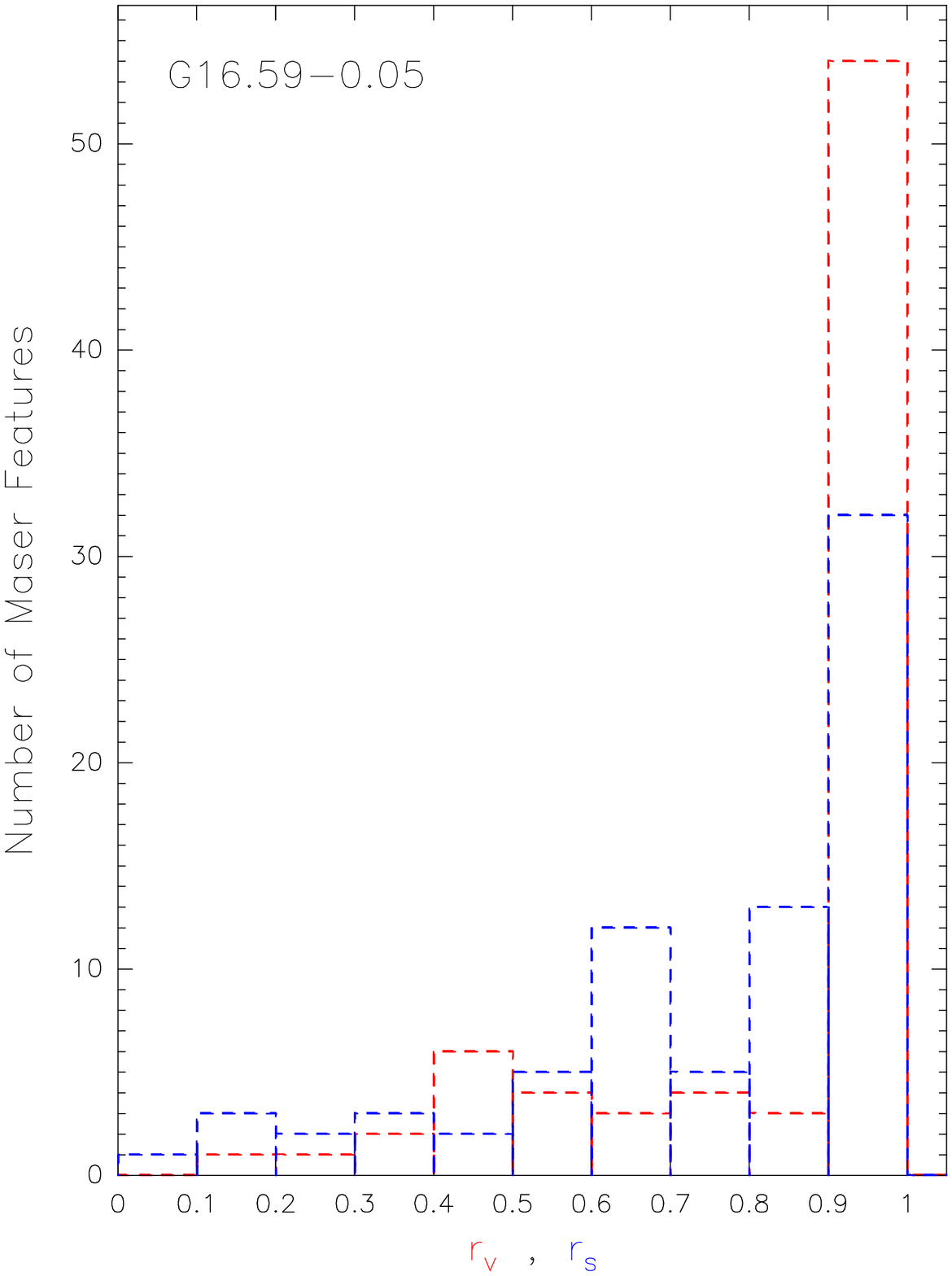}

\vspace{2cm}

\includegraphics[width=7cm]{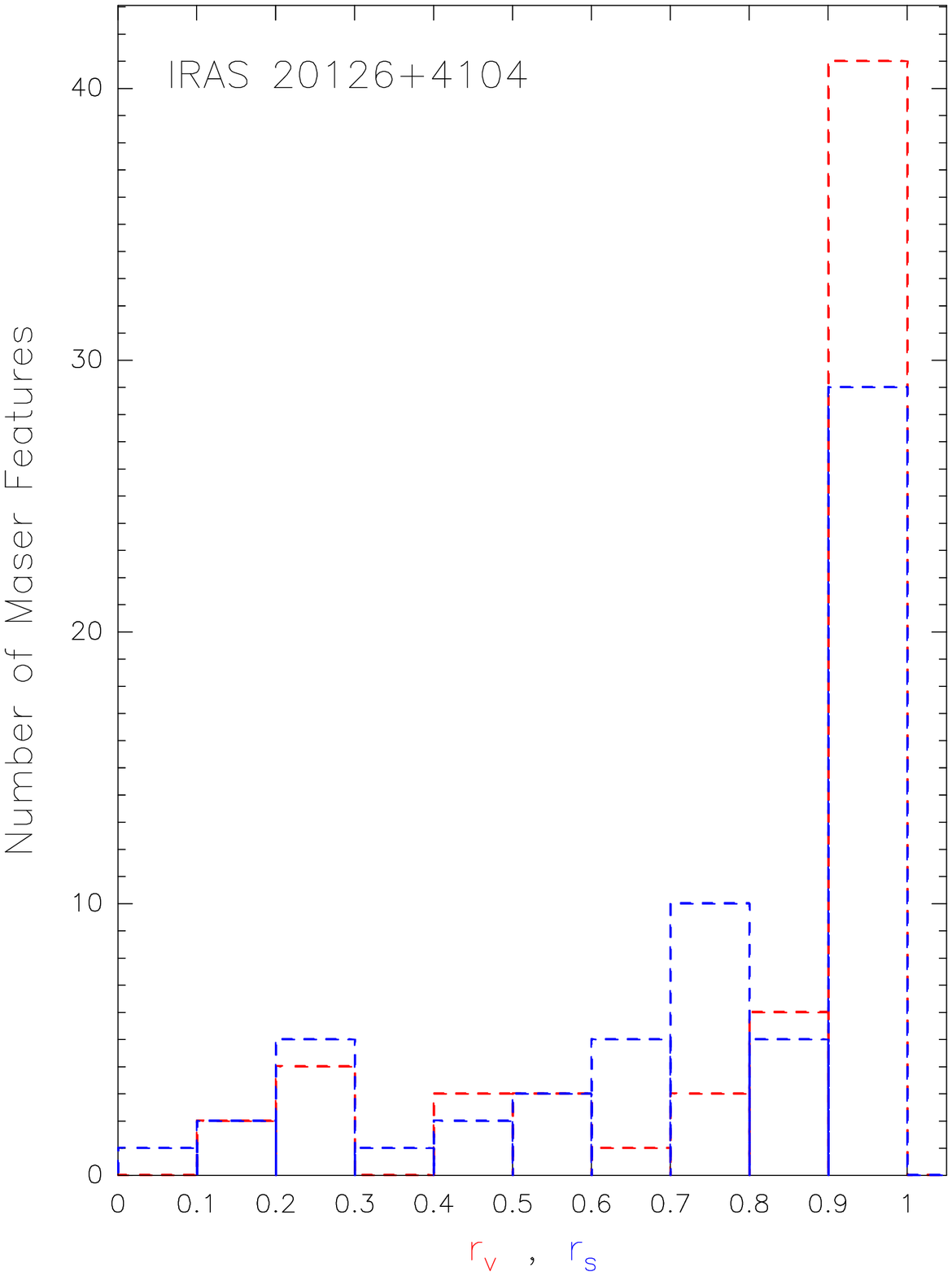}
\hspace{0.5cm}
\includegraphics[width=7cm]{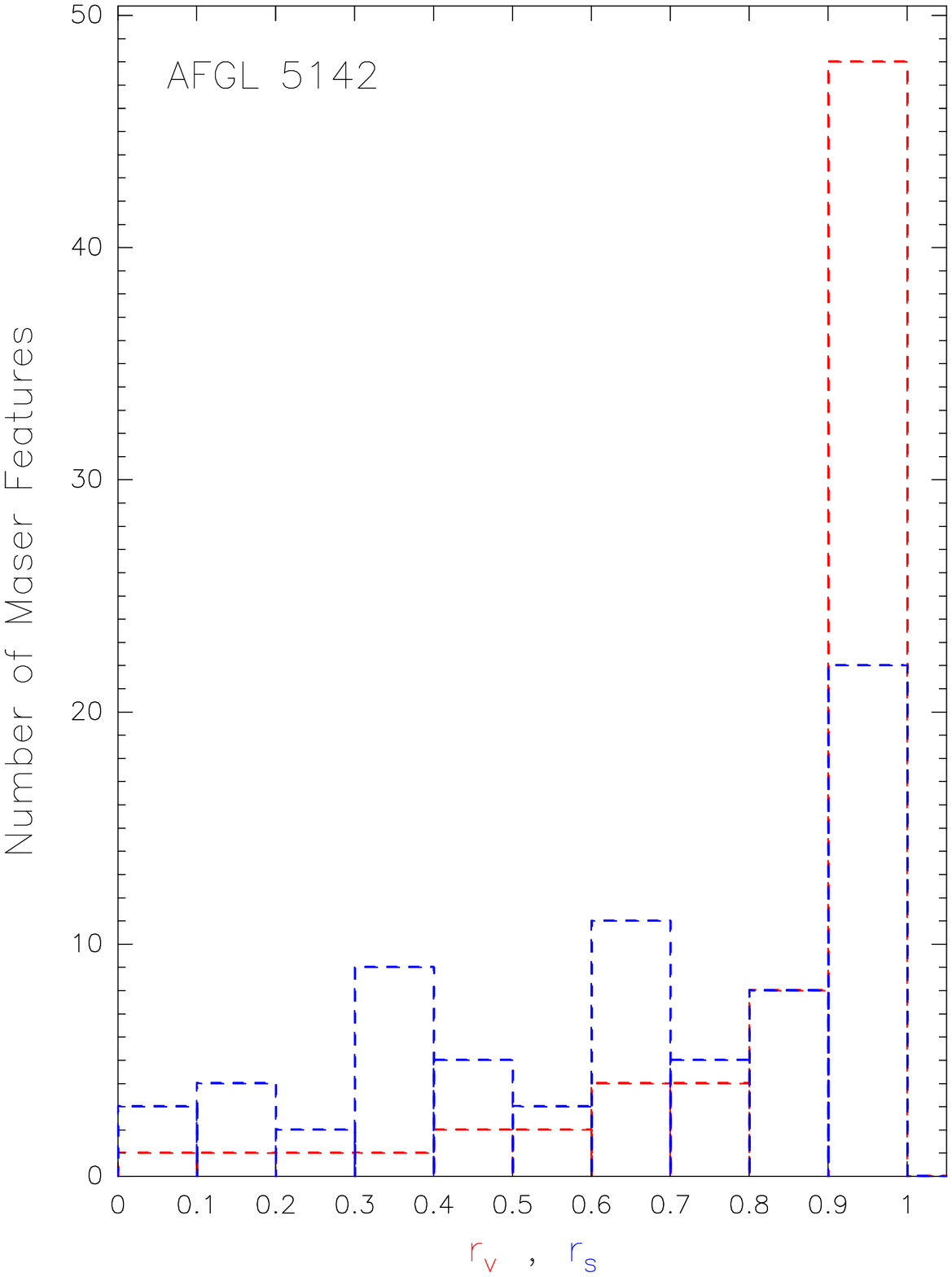}

\caption{Histograms of the linear correlation coefficients \ $r_s$ \ and  \ $r_v$.
Each panel presents the histograms of  \ $r_s$ \ ({\it blue dashed line})
and  \ $r_v$ \ ({\it red dashed line}), for one of the four maser sources,
indicated in the upper left corner of the panel. The bin size is 0.1.
}
\label{histo_rsv}
\end{figure*}

The examples shown in Fig.~\ref{feat_grad} are {\em not} rare cases
among the 6.7~GHz maser features. For each of the four target sources,
Fig.~\ref{histo_rsv} presents the histograms of the
correlation coefficients \ $r_s$ \ and  \ $r_v$ \
of the linear fits performed on all the features with three or more spots.
A large fraction (60\%--80\%) of the correlation coefficients
 showing values higher than 0.6,
 indicates that, in each source, a vast majority of maser features
has a spatial structure primarily elongated in one direction,
and also presents (at least) a hint of a regular variation in \Vlsr
along this direction.
Besides, more than half of the features present a well-defined linear structure
with quite a regular change in  \Vlsr with position, since more than half
of the derived values of \ $r_s$ \ and  \ $r_v$ \ are higher than \ 0.8.
We note that, in each source, the distribution of \ $r_v$ \
is more peaked at higher values than that of \ $r_s$. That means
that there are features showing a good correlation of \Vlsr with position
measured along the major axis of elongation, even if
their structure is only marginally elongated.
Figure~\ref{feat_grad}b illustrates this case for a representative feature.

\begin{figure*}
\includegraphics[width=10cm]{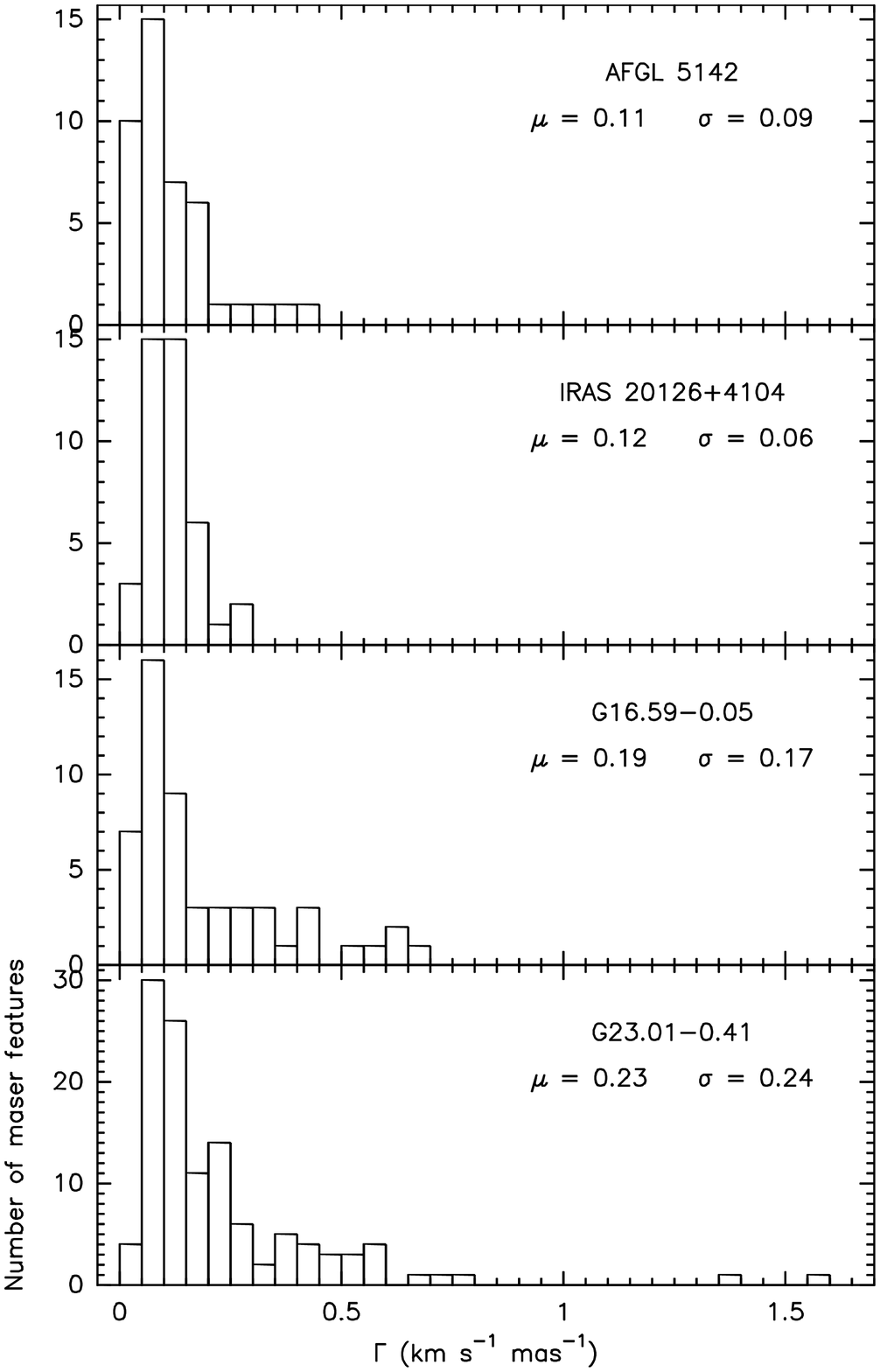}
\caption{
Histograms of  the amplitude of the \Vlsr gradient, $\Gamma$.
Each panel plots data of a single maser source, indicated in the upper right corner of the panel.
The bin size of all the histograms is \ 0.05~km~s$^{-1}$~mas$^{-1}$.
Each panel (below the source name)
also reports the arithmetic average, $\mu$, and the standard deviation, $\sigma$,
of the gradient amplitude, in units of \ [km~s$^{-1}$~mas$^{-1}$].}
\label{histo_grad}
\end{figure*}

Figure~\ref{histo_grad} presents the histograms of the amplitude of the \Vlsr gradients
for the observed  maser sources. To produce the histograms
and calculate the average and standard deviation of the distributions, we selected only
those features that presented at least some evidence of being elongated ($r_s \ge 0.6$)
and that showed a hint of a regular \Vlsr variation with position ($r_v \ge 0.6$).
The average amplitude of the \Vlsr gradients for the sources \Gp\ and \Gs\ ($\approx$0.2~km~s$^{-1}$~mas$^{-1}$)
is about twice the value derived for the sources \IR\ and \AF\ ($\approx$0.1~km~s$^{-1}$~mas$^{-1}$).

The knowledge of an accurate value of the source distance would allow us to convert
the amplitude of \Vlsr gradients from units of \ [km~s$^{-1}$~mas$^{-1}$] to units of  \ [km~s$^{-1}$~AU$^{-1}$]
and to compare the gradient amplitudes of different sources. Recently, the distance to two of the sources has been
precisely determined by measuring the parallax of the associated, strong methanol, and water masers: \IR\ is at a distance
of \ 1.64$\pm$0.05~kpc (MCR), and \Gp\ is at  \ 4.59$\pm$0.4~kpc \citep{Bru09}.
After converting to units of \ [km~s$^{-1}$~AU$^{-1}$], one finds that the distribution of gradient amplitudes in \IR\
(0.07$\pm$0.04~km~s$^{-1}$~AU$^{-1}$) overlaps with the one in \Gp\ (0.05$\pm$0.05~km~s$^{-1}$~AU$^{-1}$), although
the (proto)stars  associated to the \meth\ masers in \IR\ and \Gp\ have very different luminosities and masses (MCR,SMC2).
That lets us argue that the \Vlsr gradients internal to the 6.7~GHz masers might have a common origin in the two sources,
resulting in comparable values of gradient amplitudes. Also for the sources \Gs\ and \AF\ (using the more uncertain distance 
of 4.4~kpc and 1.8~kpc, respectively), the values of the gradient amplitude, in units of  \ [km~s$^{-1}$~AU$^{-1}$],  
are consistent with those of  \IR\ and \Gp.

\subsection{Time persistency of \Vlsr gradients}
\label{tper}

As defined in Sect.~\ref{vgra}, \Vlsr gradients are vector quantities,
oriented on the sky plane at \ P.A. = $P_s$ \ and with amplitude \ $\Gamma$.
For the two features presented in Fig.~\ref{feat_grad}, the direction on the sky plane and the amplitude
of the \Vlsr gradient appear to be fairly constant over the three observing epochs.
To investigate the variation in \Vlsr gradients with time, we selected features
persistent over two or three epochs and calculated the standard deviation of the direction
 ($\Delta~P_s$) and amplitude ($\Delta~\Gamma$) of the feature \Vlsr gradient at different epochs.
The (time-average) mean gradient is calculated by taking the mean values of the gradient components projected
along the east and the north directions. For a persistent feature, the change in time of the gradient amplitude
is measured by the ``fractional time variation"~($\Delta~\Gamma$~/~$\Gamma$), defined as the ratio
of the standard deviation, $\Delta~\Gamma$, and the mean of the values of the feature gradient amplitude, $\Gamma$,
at different epochs.

\begin{figure*}

{\large \bf (a) } \hspace{7cm} {\large \bf (b) }

\includegraphics[width=7cm]{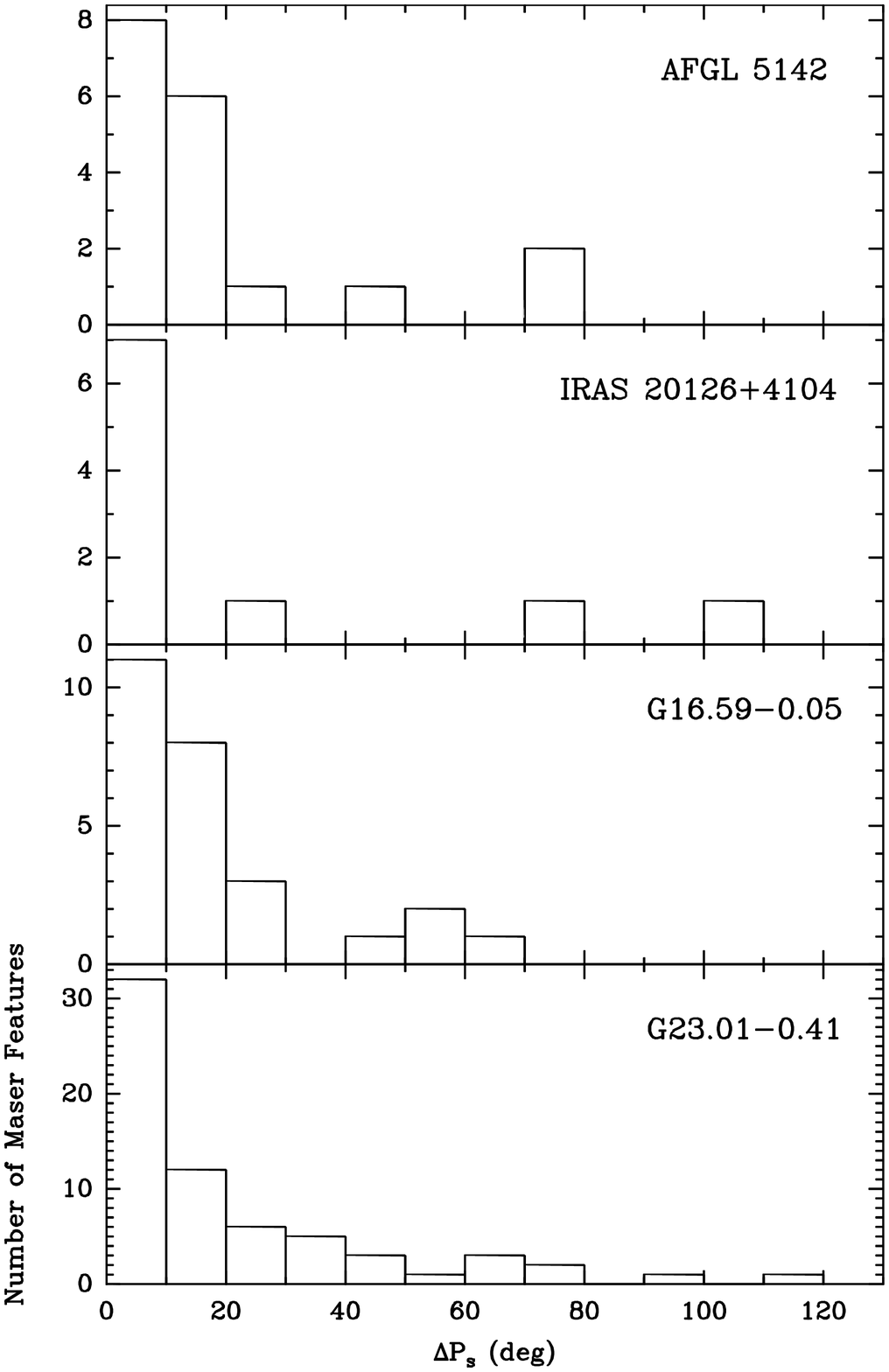}
\hspace{0.5cm}
\includegraphics[width=7cm]{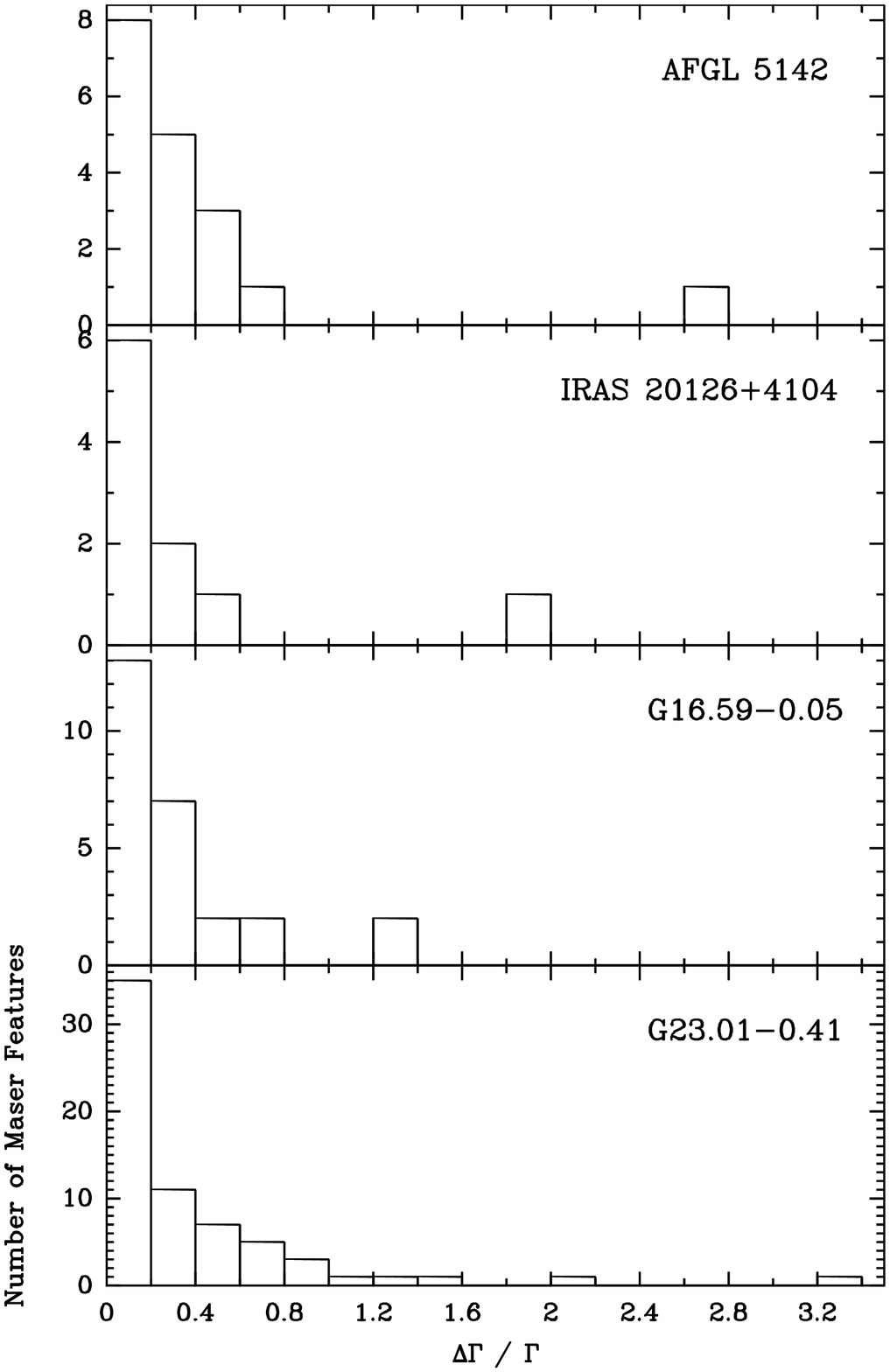}

\caption{\small
Time variation in \Vlsr gradient P.A. and amplitude. 
{\large \bf(a)}~Each of the four panels refers to the maser source indicated in the upper right corner of the panel,
and presents the histogram of the standard deviation ($\Delta~P_s$) of the values of the gradient P.A., at different epochs,
for features persisting over two or three observing epochs.
The bin size of all the histograms is \ 10$\degr$.
{\large \bf(b)}~Each of the four panels refers to the maser source indicated in the upper right corner of the panel,
and presents the histogram of the fractional time variation ($\Delta~\Gamma$~/~$\Gamma$) in the gradient amplitude,
for features persisting over two or three observing epochs.
The bin size of all the histograms is \ 0.2.
}

\label{std_gra}
\end{figure*}

Figure~\ref{std_gra} presents the histograms of \ $\Delta~P_s$ \ and \ $\Delta~\Gamma$~/~$\Gamma$
for each of the four maser sources.
In each source, on a timescale of several years, the amplitude of the \Vlsr gradient for most of the persistent features
changes less than \ 20--40\%, and the variation in the gradient orientation is generally less than \ 20$\degr$.
Measurement errors can contribute to enlarge the calculated spreads in time,
so that the effective time variation could be even smaller than what is shown in Fig.~\ref{std_gra}.
Therefore, we can conclude that most of the 6.7~GHz maser internal \Vlsr gradients
appear to be remarkably stable in time on a timescale of (at least) several years.

\subsection{Ordered spatial and angular distribution of \Vlsr gradient directions}
\label{spor}

\begin{figure*}
{\large {\bf(a)} \Gp}

\includegraphics[width=7cm,angle=-90]{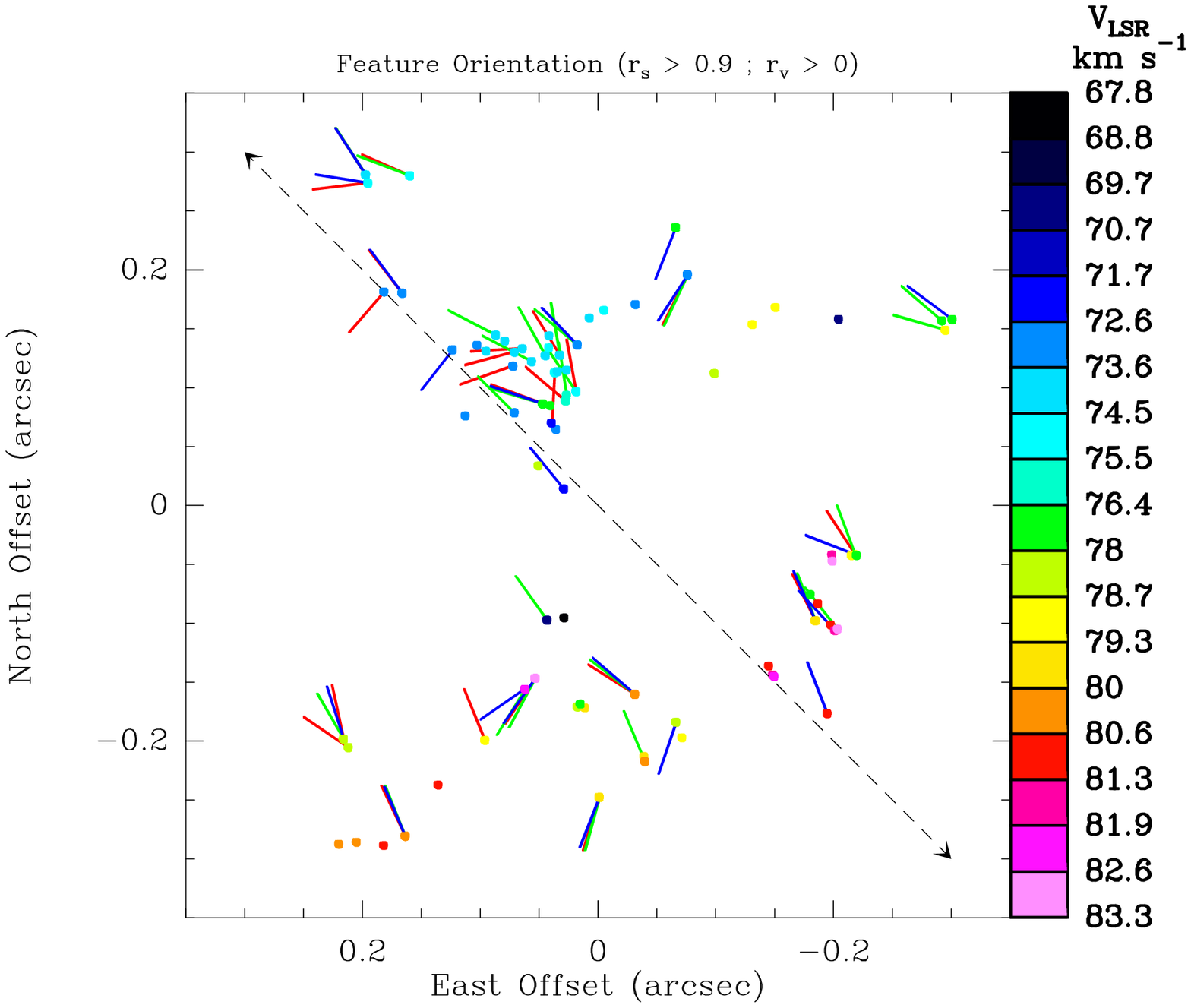}
\hspace{0.5cm}
\includegraphics[width=7cm,angle=-90]{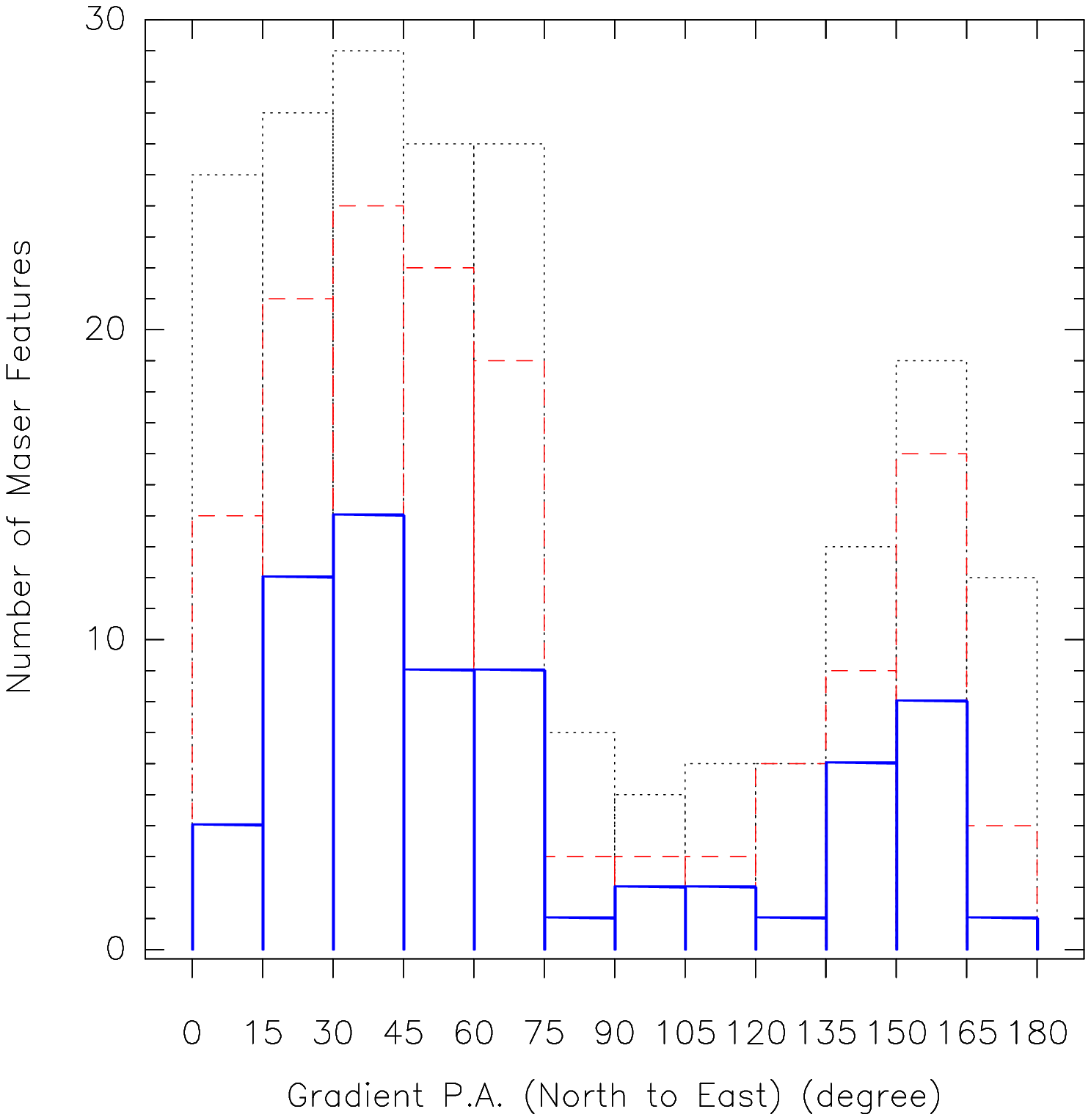}

\vspace{1cm}

{\large {\bf(b)} \Gs}

\includegraphics[width=6.5cm,angle=-90]{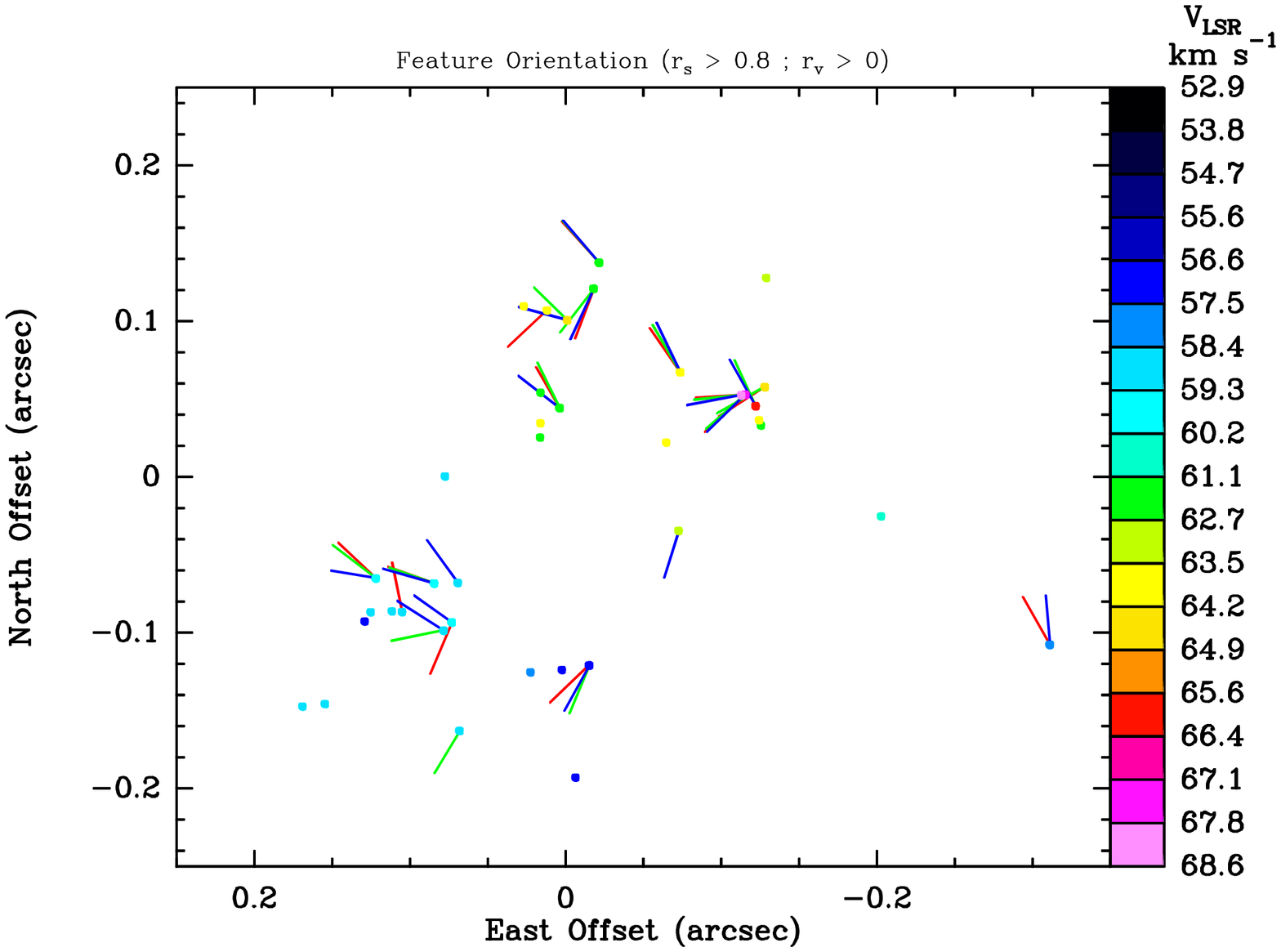}
\hspace{0.5cm}
\includegraphics[width=7cm,angle=-90]{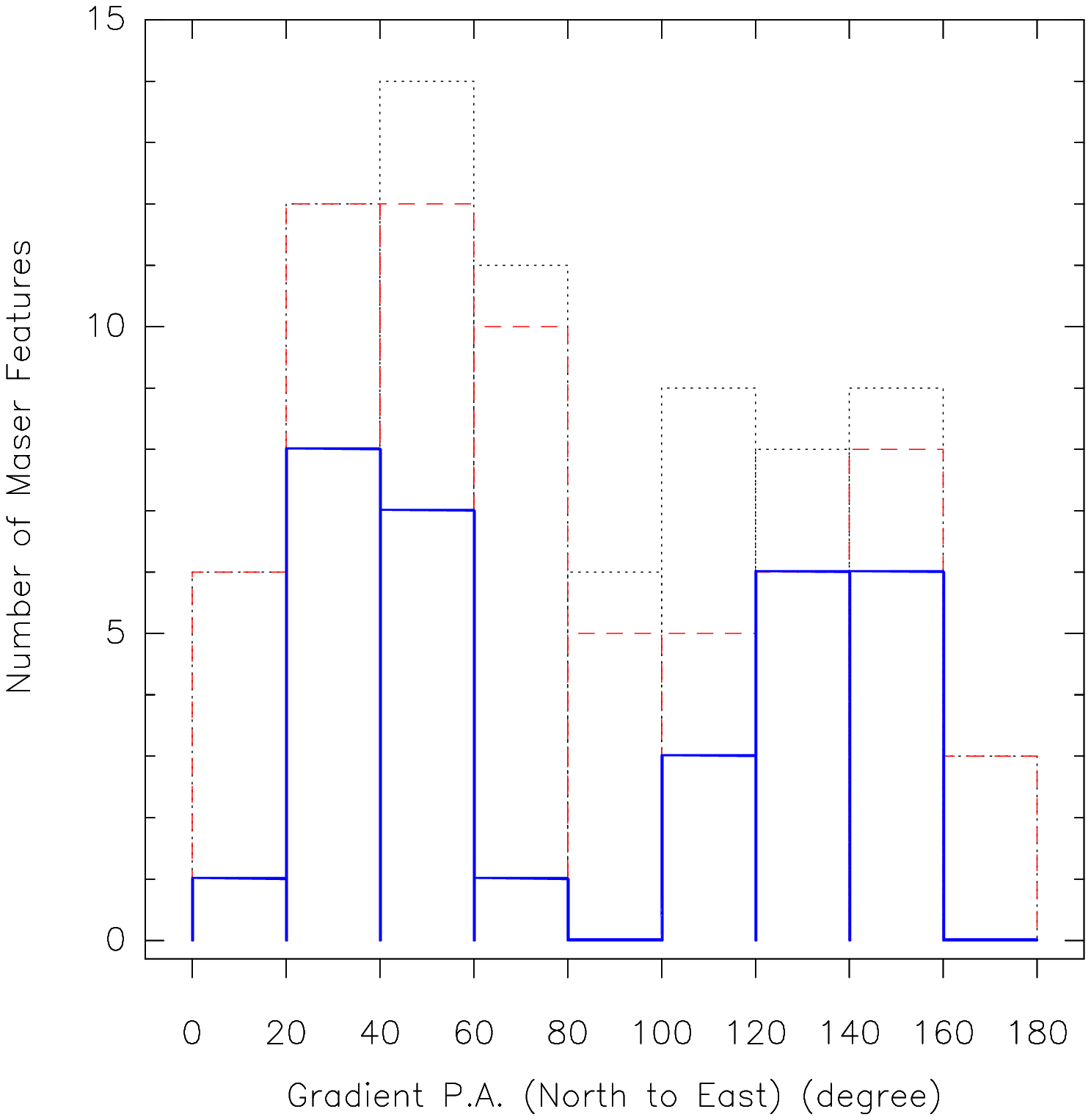}

\vspace{-0.1cm} \caption{\tiny The panels in each row present the spatial distribution and the histogram 
of the \Vlsr gradient directions for the maser source indicated on the left side above the panels.
{\bf Left Panel:} Spatial distribution of the \Vlsr gradient directions. {\it Colored dots} report the position
of the maser features detected over the three observing epochs, with colors indicating the feature $V_{\rm LSR}$.
The velocity-color conversion code is shown in the wedge on the right side of the panel, with {\it green}
denoting the systemic \Vlsr of the maser source. {\it Colored segments}
associated to maser features give the direction of the feature \Vlsr gradient  (with P.A. varying in the range \ $0\degr$--$180\degr$),
with different colors to distinguish the observing epoch: red, green, and blue, for the first, second, and third epochs, respectively.
The segment length is proportional to the value of the correlation coefficient,$r_s$, of the linear fit to the spot positions,
from which the gradient direction is derived.
The plot reports only the feature gradients with better defined directions, corresponding to more linear spot distributions
and higher values of the correlation coefficient, $r_s$:  \ $r_s > 0.9$ \ for \Gp, and \ $r_s > 0.8$ \ for \Gs.
For the sources \Gs\ and \Gp, feature positions are relative to the maser's ``center of motion", as defined in
SMC1 and SMC2, respectively. In the plot of \Gp, the {\it dashed arrow}  indicates the direction of the collimated jet
traced close to the (proto)star by the 22~GHz water masers (SMC2).
{\bf Right panel:} Histograms of the P.A. of the feature gradient directions.
{\it Dotted black}, {\it dashed red}, and {\it continuous blue} lines show the histograms
of the gradient P.A. for features with increasingly better linear structure, corresponding to values of the correlation coefficient
\ $r_s > 0 $, $r_s > 0.5$, and \ $r_s > 0.9$, respectively. The histogram bin size is \ 15$\degr$
for the source \Gp, and \ 20$\degr$ for the source \Gs.}

\label{sky_dis}
\end{figure*}

\addtocounter{figure}{-1}

\begin{figure*}

{\large {\bf(c)} \IR}

\includegraphics[width=6cm,angle=-90]{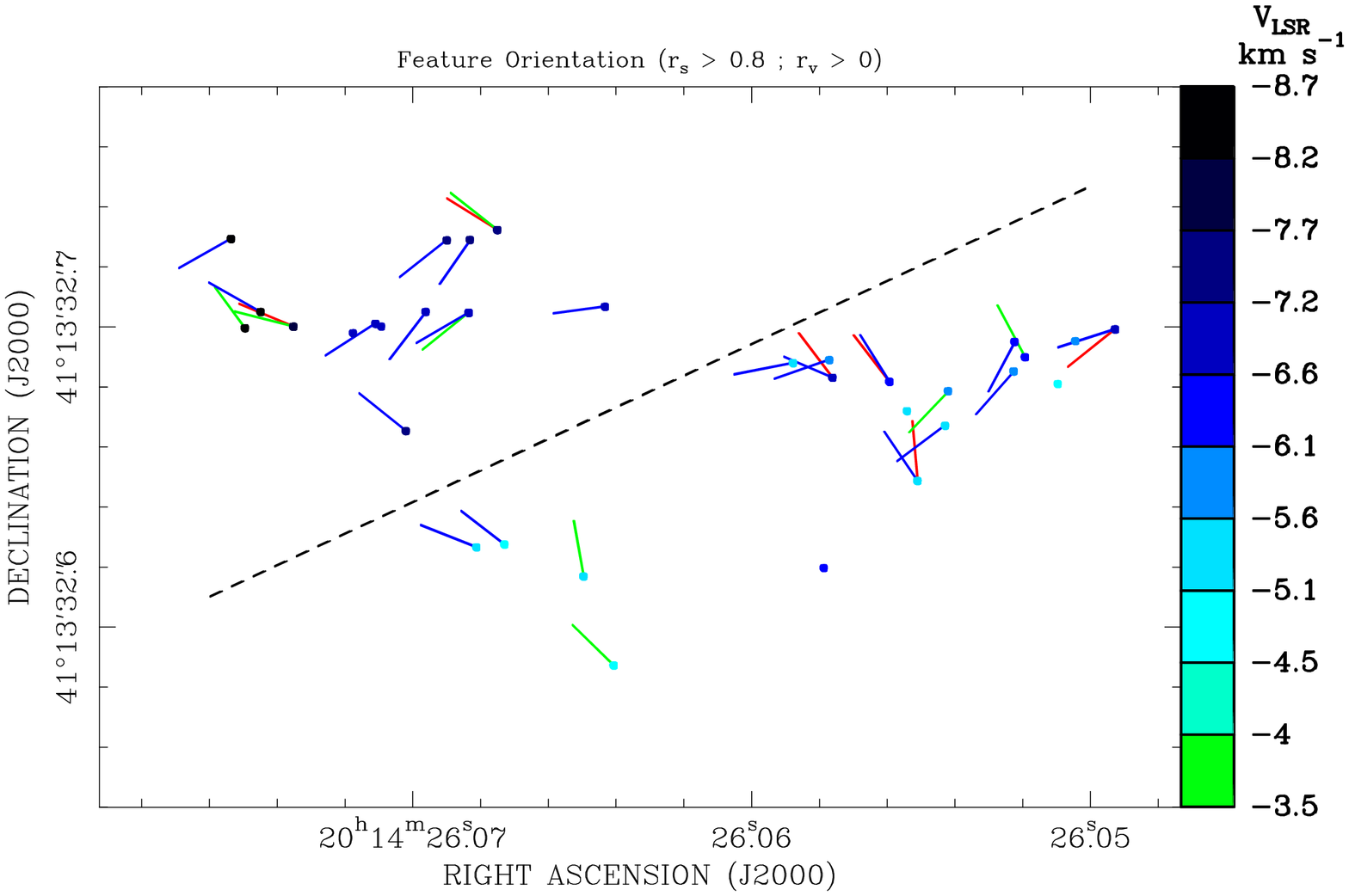}
\hspace{0.5cm}
\includegraphics[width=7cm,angle=-90]{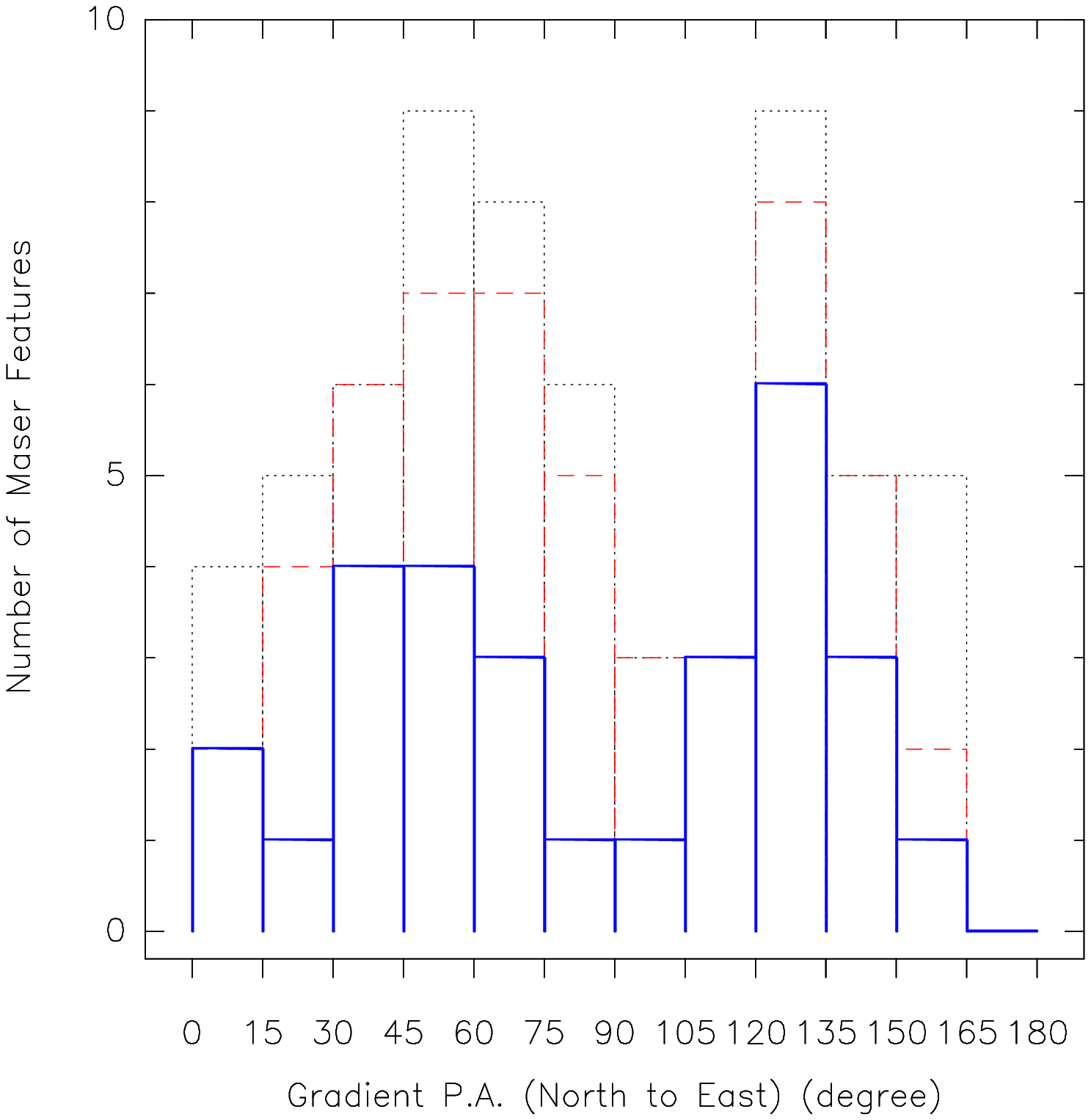}

\vspace{1cm}

{\large {\bf(d)} \AF}


\includegraphics[width=7cm,angle=-90]{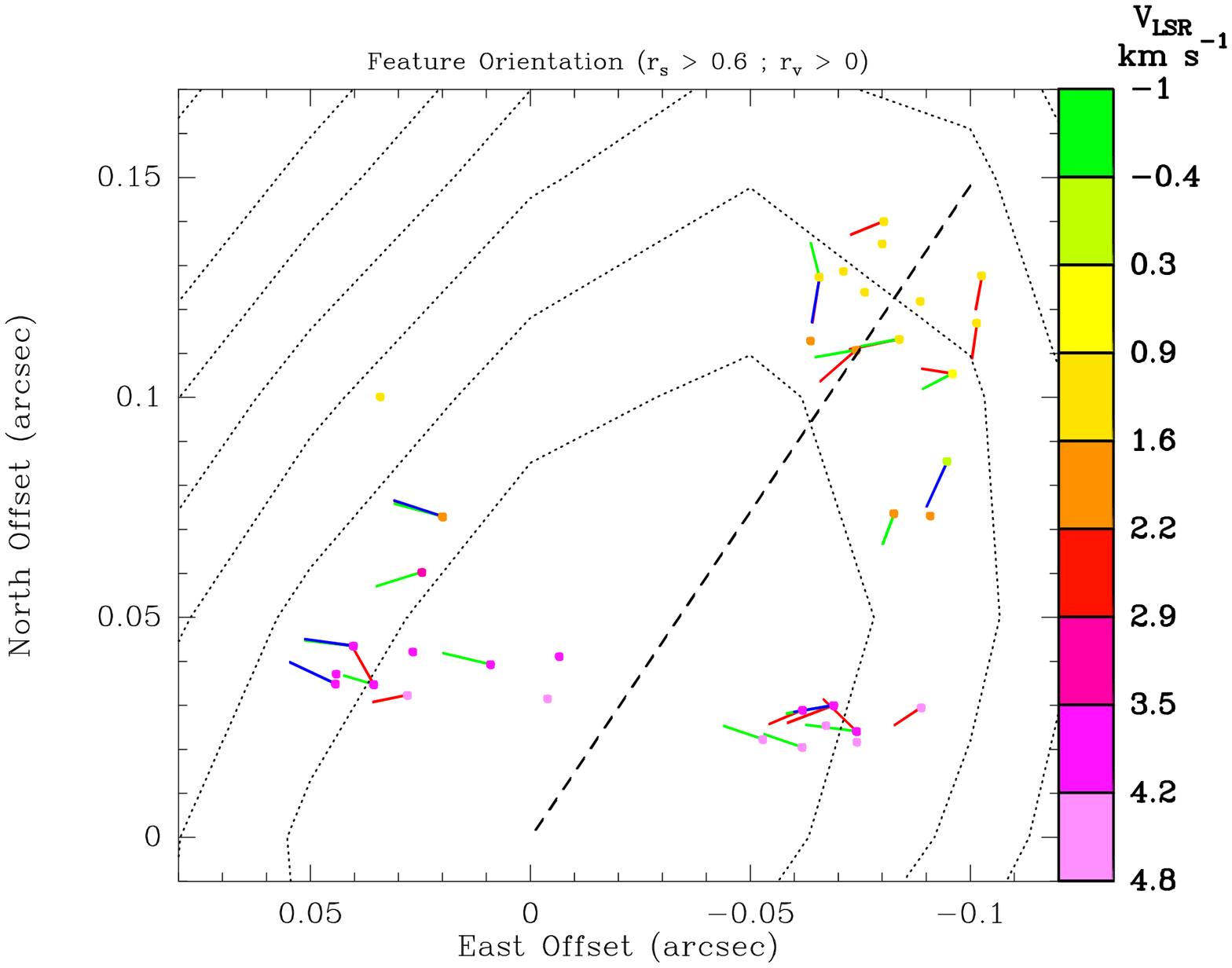}
\hspace{0.5cm}
\includegraphics[width=7cm,angle=-90]{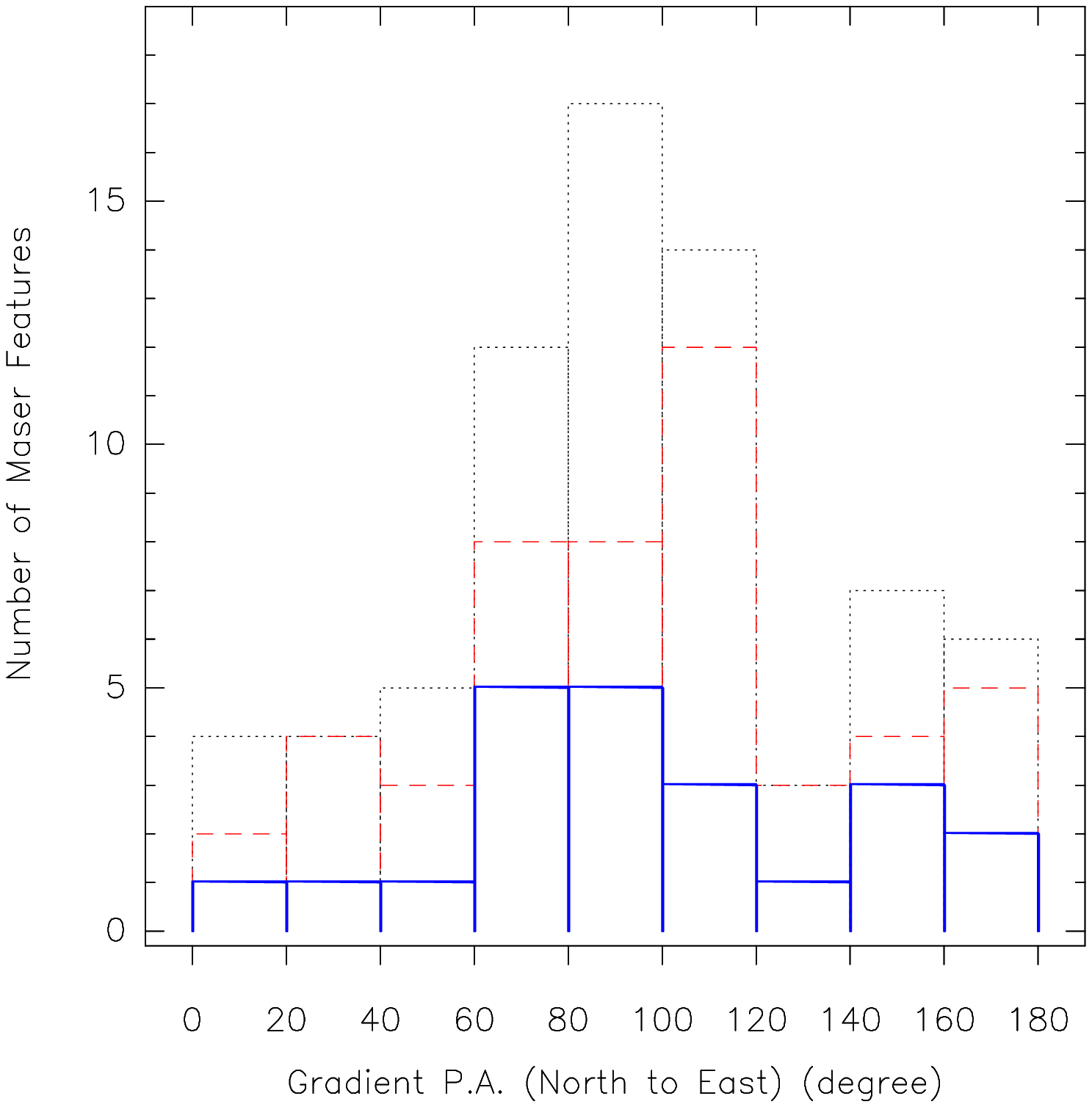}

\caption{\tiny The panels in each row present the spatial distribution and the histogram 
of the \Vlsr gradient directions for the maser source indicated on the left side above the panels. 
{\bf Left panel:} Spatial distribution of the \Vlsr gradient directions.
{\it Colored dots} and {\it colored segments} have the same meaning as
in Figs.~\ref{sky_dis}a~and~\ref{sky_dis}b. For the source \IR, absolute maser positions are reported,
relative to the observing epoch 2004 November 6 (MCR, Fig.~2). In the source \AF,  feature positions
are relative to the peak of the VLA 1.3~cm continuum observed toward the 6.7~GHz masers
\citep[see Fig.~2]{God07}, whose contour map is shown by a {\it dotted line}. In both the plots of  \IR\ and \AF,
the {\it dashed line} gives the direction of the collimated jet traced close to the (proto)star by the 22~GHz water masers.
The plot reports only the feature gradients with better defined directions, corresponding to more linear spot distributions
and higher values of the correlation coefficient,  $r_s$:  \ $r_s > 0.8$ \ for \IR, and \ $r_s > 0.6$ \ for \AF.
{\bf Right panel:} Histograms of the P.A. of the feature gradient directions.
{\it Dotted black}, {\it dashed red}, and {\it continuous blue} lines have
the same meaning as in Figs.~\ref{sky_dis}a~and~\ref{sky_dis}b.
The histogram bin size is \ 15$\degr$ for the source \IR, and \ 20$\degr$ for the source \AF.
}
\end{figure*}

For each of the four maser sources, Figure~\ref{sky_dis} shows the distribution on the plane of the sky
of the directions of \Vlsr gradients and the histogram of the P.A. of gradient directions.
To produce these plots, we neglected the sign of the  \Vlsr gradient, and the P.A.
of negative gradients ($180\degr \le P_s \le 360\degr$) was folded
into the range \ $0\degr$ -- $180\degr$.

Looking at the histograms of the gradient P.A., it is evident that, in each of the four sources,
the directions of feature gradients do not distribute uniformly, but tend to concentrate into
specific P.A. ranges. For the sources \Gp, \Gs, and \IR, the P.A. distribution presents
two broad, but clearly distinct peaks at the angles \ $40\degr$$\pm$$20\degr$ and \ $140\degr$$\pm$$20\degr$.
For the source \AF, only a single, broad peak, over the angles \ $90\degr$$\pm$$30\degr$, is clearly
visible. For the sources \Gp, \Gs, and \IR, with a larger number of detected maser features and better statistics,
the peaks in the distribution of gradient P.A. become narrower when selecting gradients with increasingly better
measured directions (i.e. corresponding to linear fits with increasingly higher values of the correlation coefficient, $r_s$).
That reinforces the impression that the two-peak distribution has a physical meaning, and it does not result from a
bias in our data analysis. 

The plots of the spatial distribution of the \Vlsr gradient directions illustrate that,
in the sources \Gp, \Gs, and \IR, the two groups of features with gradients
approximately perpendicular to each other, are not confined in two different regions
of the sky, but are spread over a similar area and are observed to almost overlap
(with separation of only $\approx$10~mas) along some lines-of-sight. In these three
maser sources, the two groups of gradients also have a similar spatial distribution
of \ $V_{\rm LSR}$. 
The case of the source \AF\ may be different. In this source, the more red-shifted
features, confined to the south and southeast of the whole maser distribution, have
\Vlsr gradients directed close to east-west (P.A. $\approx$ 90$\degr$), whereas most of the
"yellow" features, placed to the north and northwest,  show gradients directed close to
north-south (P.A. $\approx$ 0$\degr$ or 180$\degr$). For this source, the poor
statistics of well-measured gradient orientations very likely prevents us from clearly detecting a second peak
in the distribution of gradient P.A. close to  \ 0$\degr$  and/or \ 180$\degr$. In any case,
in \AF, differently from the other three sources,
the direction of the maser \Vlsr gradient appears to be correlated with the feature position.

Section~\ref{dis} discusses the peculiarities observed in the spatial and angular distribution
of the 6.7~GHz maser \Vlsr gradients, and proposes a simple kinematical interpretation.

\section{Comparison with 6.7~GHz maser proper motions}
\label{gra_pro}

The time persistency of most of the 6.7~GHz maser \Vlsr gradients on timescales of (at least) several years,
and the regular (spatial and angular) distribution of the gradient directions observed towards all the four maser targets
suggests that the maser internal gradients
might reflect the ordered gas motions observed over the whole maser region (size of $\sim$100--1000~AU)
on small (linear) scales ($\sim$10~AU). In all sources but \Gs, our 22~GHz water maser observations show
 a fast wide-angle wind and/or a collimated jet emerging from the same (proto)star exciting
the methanol masers. 
Over the whole region, the 6.7~GHz masers present
a rather ordered 3-D~kinematics, suggesting either rotation (in \Gs), a combination of rotation plus expansion
(\Gp\ and \IR), or infall (\AF). 

\begin{figure*}
\centering
\includegraphics[width=8cm, angle=-90.0]{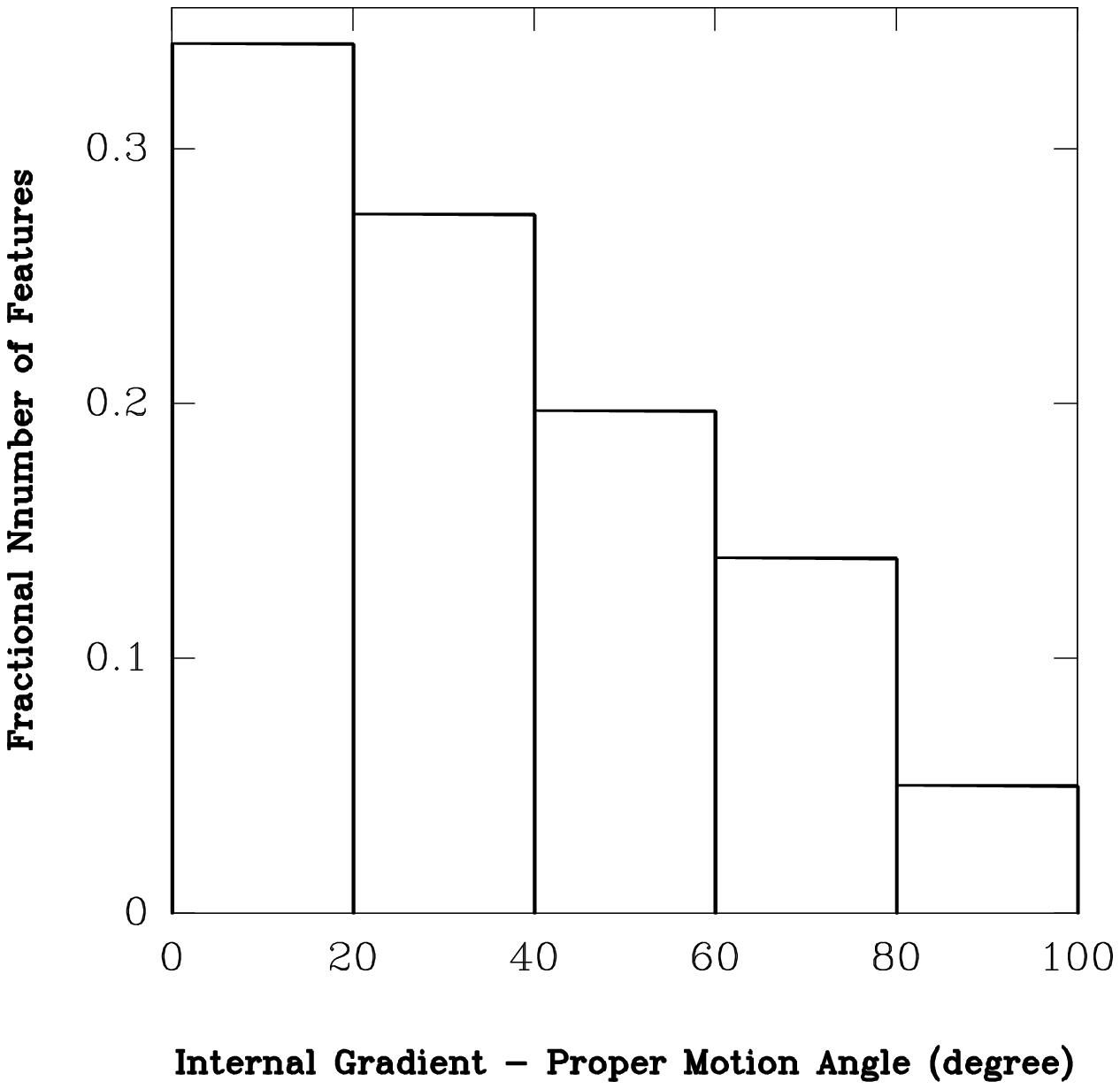}
\caption{
Histogram of the distribution of the angle between the
direction of the \Vlsr gradient of a maser feature
and its proper motion vector. The histogram bin size is \ 20$\degr$.
The histogram reports
the fractional number of maser features 
 averaged over the four target sources.
}
\label{gra_vel}
\end{figure*}

If the \Vlsr gradients internal to the 6.7~GHz masers reflect ordered, large-scale gas motions,
one would expect them to be related to the gas velocities.
For instance, should the \Vlsr gradients be produced in either a rotating structure or a Hubble outflow,
one would expect the directions of the feature gradient and proper motion to project at a close angle on the sky.
Figure~\ref{gra_vel} presents the source-average distribution of the angle between the
direction of the \Vlsr gradient of a maser feature and its proper motion vector.
The 6.7~GHz maser proper motions are read from Table~3 of SMC1 for \Gs,
 from Table~4 of SMC2 for \Gp, from Table~1 of MCR for \IR, and from GMS2 for \AF.
The acute angle between the feature gradient and proper motion directions
on the sky plane is used to measure the angular separation.
To produce the plot in Fig.~\ref{gra_vel}, feature gradients with loosely defined directions
(i.e. with \ $r_s \le 0.5$~and~$r_v \le 0.5$)
 were discarded, and, for persistent features, the mean gradient averaged over the three epochs was calculated.
Features that have both the \Vlsr gradient and the proper motion reliably measured,
are a few tens for the sources \Gp\ (36) and \AF\ (20), but
significantly less for \Gs\ (8) and \IR\ (7). To increase the statistics,
we calculated  the distribution of the {\em fractional} number of features for each source
(dividing the number of features of a histogram bin by the total number of features),
and then averaged over the four sources.

Figure~\ref{gra_vel} provides some evidence that the \Vlsr gradients tend
to be oriented at relatively small angles ($\le40\degr$)
from the direction of the maser proper motions. The actual
distribution might be significantly more peaked at small angles
than observed, since the uncertainty
in the measurements of the gradient and proper motion orientations
can be large ($\ge$20$\degr$), and can contribute significantly to broaden
the peak. The basic characteristics of the source-average plot
of Fig.~\ref{gra_vel} do not result from
one source's distribution dominating the other ones,
but, conversely, in all four sources there is a similar
tendency towards small angular separations between
the maser gradients and proper motions. 
Therefore we can conclude that in each of the
observed sources, the data are consistent with having the \Vlsr gradients and 
proper motion vectors in the same direction on the sky, 
considered the measurement uncertainties.

\citet{Fis06a}, studying the 1.6~GHz OH masers towards the W3(OH) UC~H{\sc ii} region,
did not note any correlation between the orientation of the \Vlsr
gradients and the proper motions of the OH masers (measured by \citealt{Blo92}).
Over the whole extent of the UC~H{\sc ii} region ($\approx$3000~AU in size),
\citet{Fis06a} and \citet{Fis07a} did not find any large-scale organization
in the distribution of the OH \Vlsr gradients, and then suggest that the observed gradients
should trace local phenomena, which are possibly associated to AU-scale turbulent fluctuations
in the sky-projected velocity field. However, the environment traced by the 6.7~GHz masers
in each of our target sources appears to be different from that of an UC~H{\sc ii} region.
The typical 6.7~GHz maser velocities (5--10~\kms) are significantly higher than those of the
OH masers in W3(OH) (a few \kms). Whereas in W3(OH) and in other sources as well (\Gp; see SMC2),
the OH masers appear to mark gas at greater distance ($\ge10^3$--10$^4$~AU) from the (proto)star and
to trace  the slow expansion of the UC~H{\sc ii} region, the 6.7~GHz masers
are more likely associated with gas accreting, rotating, and outflowing,
in the proximity (within hundreds of AU) to the (proto)star.
Differently from the OH, we believe that the measured gradients of 
CH$_3$OH masers trace gas kinematics on AU-scales, as discussed in detail in the next section.

\section{Nature of the 6.7~GHz maser \Vlsr gradients}
\label{dis}

In the following, we discuss the hypothesis that the 6.7~GHz maser \Vlsr gradients trace, on very small scale ($\lesssim$10~AU), 
the bulk-ordered motion measured  on large scales from maser kinematics: outflow or infall (Sect. 5.1) and rotation (Sect. 5.2) in high-mass (proto)stars.

\subsection{Do methanol maser \Vlsr gradients trace outflow/infall?}
\label{outflow}


We consider first the case that the 6.7~GHz maser \Vlsr gradients are associated with the large-scale outflows traced by the water masers in the sources \Gp, \IR\ and \AF.

In the source \IR, the 3-D~velocity~field of the 22~GHz water masers has been successfully reproduced
with a model of a conical, Hubble flow, with maser velocities increasing linearly with the distance from the star.
Assuming that the water maser velocity distribution can also be described in terms of an Hubble flow in the other sources,
from the measured 3-D~velocities it is possible to derive a typical velocity gradient in the outflowing gas.
Considering that the observed water masers generally move close to the plane of the sky, one has to
divide the maximum observed value of {\it transverse} velocities by the sky-projected semi-length of the outflow.
This calculation gives a gradient of \ 0.2~km~s$^{-1}$~mas$^{-1}$ \ and \ 0.15~km~s$^{-1}$~mas$^{-1}$ \ for the jets
observed in the sources \AF\ and \IR, respectively. For the source \Gp, we obtain a gradient of   \ 0.5~km~s$^{-1}$~mas$^{-1}$ \
for the fast and compact wide-angle wind located close to the (proto)star and a much lower value of  \ 0.06~km~s$^{-1}$~mas$^{-1}$ \
for the collimated jet traced by two groups of water masers at a greater distance from the (proto)star.
In each of the three sources, the gradient in the outflow velocities derived from the water masers
is an upper limit to the average  value of the 6.7~GHz maser \Vlsr gradients (see Fig.~\ref{histo_grad}).
That is consistent with the fact that 6.7~GHz masers are found to move much more slowly than water masers do.
One possible interpretation is that water masers trace fast, shocked gas closer to the jet axis, 
while methanol masers might originate in relatively slow-moving material, entrained in the outflow at a larger distance from the jet axis.
Looking at specific cases, in the source \IR\ MCR show that the entraining process takes place for a group of 6.7~GHz features
with proper motions directed close to the jet direction, while for the source 
\AF\ \ we propose that the 6.7~GHz maser 3-D~velocities trace infall rather than outflow. Regardless of the specific source case, 
this simple discussion shows that, in each of the three sources where an outflow or infall is detected, the gas moves fast enough 
{\it to be able} to account for the \Vlsr gradients internal to the 6.7~GHz masers.

Figure~\ref{sky_dis} shows that  the directions of the \Vlsr gradients in three sources
tend to cluster in two P.A. ranges (separated by about \ 90$\degr$--100$\degr$), while a single peak
in the distribution of gradient P.A. is evident for the source \AF. For the three sources with an observed
water maser jet, the orientation of the jet axis is reported in the plot of the
spatial distribution of gradient directions. It is interesting to note that,
for both sources \Gp\ and \IR, the P.A. of the jet axis (about \ $50\degr$ \ and  \ $115\degr$ \
for the first and second sources, respectively) is close to one of the two peaks
in the distribution of gradient P.A.. We consider that,
while the direction of the jet axis is the one along which \Vlsr gradients associated with the
outflow should project, the direction perpendicular to the jet
should be the preferred orientation of gradients due to (almost edge-on) rotation about the jet.
This qualitative argument can simply explain the presence of the two peaks in the distribution of gradient
P.A. for the sources  \Gp\ and \IR, and lends further support to the idea of a correlation between
the small-scale \Vlsr gradients inside the 6.7~GHz maser features and the ordered gas motions observed over
the whole maser region.

\subsection{Do methanol maser \Vlsr gradients trace rotation?}
\label{disk}

One of  the simplest interpretations of a regular change in \Vlsr with position across the plane of the sky
is in terms of rotation. If the spots of a
6.7~GHz maser feature traced a portion of an annulus of a rotating sphere of gas,
seen almost edge-on, that could explain the (sky-projected) linear distribution of spots
and also the linear variation in \Vlsr with (sky-projected) spot positions. Assuming that
the rotation is gravitationally supported, one can easily derive a relation between
the average gas (mass) density, $\overline{\rho}$, and the measured gradient amplitude,  $\Gamma$:
\begin{equation}
\label{den_gra}
 \Gamma^2 = G \, \frac{M_s}{R{_s}{^3}} = 4/3 \, \pi \, G \, \overline{\rho}
\end{equation}
where \ $G$ \ is the gravitational constant, and \ $ M_s$ and $ R_s$ are the mass and the radius
of the sphere, respectively.
Using an average value of \ $\Gamma = 0.05$~km~s$^{-1}$~AU$^{-1}$
(as derived above for the sources \IR\ and \Gp), one finds \ $\overline{\rho} = 4 \, 10^{-13}$~g~cm$^{-3}$,
corresponding to a number density of hydrogen molecules \ $n_{H_2} = 10^{11}$~cm$^{-3}$.
Using this average density, indicating with \ $R_{10}$ \ the radius of the maser annulus in units of \ 10~AU,
the mass of the gas sphere inside the annulus radius is \ $ M_s = 2.7 \, 10^{-3} \, R_{10}^{3}$~M$_{\odot}$.

The obtained value of  \ $\overline{\rho} $ \ and the expression for \ $M_s$  \ can be used to further constrain
the properties of the putative rotation traced by the 6.7~GHz maser \Vlsr gradients.
Excitation models of the 6.7~GHz methanol masers \citep{Cra05} predict that this maser emission
should be quenched for \ $n_{H_2} \ge 10^9$~cm$^{-3}$.
Since this value is  two orders of magnitude lower than the estimate
of the average density \ $\overline{\rho}$ \ of the masing sphere,
we can conclude that it is unlikely that 6.7~GHz maser features trace a portion of an annulus
of a rotating sphere of gas, unless the gas mass is very concentrated close to the center. Then,
if the \Vlsr gradients are tracing rotation,  Keplerian rotation should be preferred over
solid-body rotation, which results from an homogenous mass distribution.
The typical (sky-projected) size of a feature's spot distribution (see examples reported in Fig.~\ref{feat_grad})
ranges from a few mas to 10~mas, or  \ 5--50~AU, for a source distance from a few~kpc up to 5~kpc.
If the diameter of the maser annulus were comparable to the observed feature size,
the mass \ $ M_s$ \ enclosed inside the annulus radius, would be less or much less than a few percent of a solar mass.
In each of the four sources, 6.7~GHz maser emission is observed close to a massive proto(star)
where the density and the temperature of the gas are relatively high ($n_{H_2} \approx 10^6$~cm$^{-3}$, T$\approx$100--200~K).
Since the characteristic Jeans mass of such a dense and warm gas should be about 1~M$_{\odot}$,
it is very unlikely that most of maser features in the observed sources
trace a self-gravitating body with a mass of thousandths or hundredths of a solar mass.
Therefore we favor the alternative view where the annulus radius is significantly larger than the feature size,
so that the enclosed mass \ $ M_s$ \ would be comparable to or higher than 1~M$_{\odot}$.
Our conclusion is that, if the gradients internal to the 6.7~GHz masers traced rotation,
a maser feature (with size $\sim$10~AU) would more likely trace only a {\em small arc}
of a ring of matter in {\em Keplerian} rotation around a central mass \ $ \gtrapprox 1$~M$_{\odot}$.

In  Section~\ref{qua_dis}, we discuss the specific case of \Gs, 
where the methanol maser \Vlsr gradients, under the assumption of Keplerian rotation, 
 enabled us to constrain the (proto)star position and mass, finding good agreement with the values
derived from the (3-D) maser spatial and velocity distribution.

\subsubsection{\Vlsr gradients tracing rotation in \Gs}
\label{qua_dis}


Among the four maser targets, the source \Gs\ is perhaps the one where the 3-D~velocity~field
of the 6.7~GHz methanol masers traces the simplest kinematics, consisting in
a rotating structure, elongated \ 0$\farcs$3--0$\farcs$4 \ towards the southeast-northwest direction
and  inclined \ $\approx$30$\degr$ \ with the plane of the sky (SMC1, Fig.7).
No collimated water maser jet is observed in this source, and it is likely that most of the 6.7~GHz
\Vlsr gradients trace rotation, since the maser internal gradients are mainly directed transversally
to the line connecting the maser position with the center of rotation (see Fig.~\ref{sky_dis}b).
Now we apply the general arguments of Sect.~\ref{disk} to this specific source,
making use of the maser \Vlsr gradients to constrain the (proto)star position and mass.
Equation~\ref{den_gra} indicates that, if maser internal gradients trace Keplerian rotation around a
central mass $M_s$, the product of the cube of the maser distance to the (proto)star, $R_{s}^3$,
and the square of the gradient amplitude, $\Gamma^2$, is a constant proportional to the stellar mass.
Equation~\ref{den_gra} is strictly valid for edge-on rotation. More in general,
if the rotation axis is inclined by an angle \ $i$ \ with the line-of-sight, for sufficiently low
values of the polar angle\footnote{The position of the orbiting maser feature is described using polar
coordinates centered on the (proto)star, with the polar angle $\theta$ taken to be zero when the polar radius is at
the minimal separation ($90-i$) from the line-of-sight.} $\theta$, where \mbox{$0 \le \theta \le \arctan(\cos^{-1}(i))$},
the same relation holds replacing the central mass \ $M_s$ with the product \ $M_s \, \sin^2(i)$. The maser gradients in \Gs,
tracing a rotating structure with an inclination angle \ $i\approx$30$\degr$, should satisfy a relation equivalent
to Eq.~\ref{den_gra} over a wide range of polar angles \ ($0 \le \theta \le 50\degr$).

\begin{figure*}
\includegraphics[width=6.3cm, angle=-90.0]{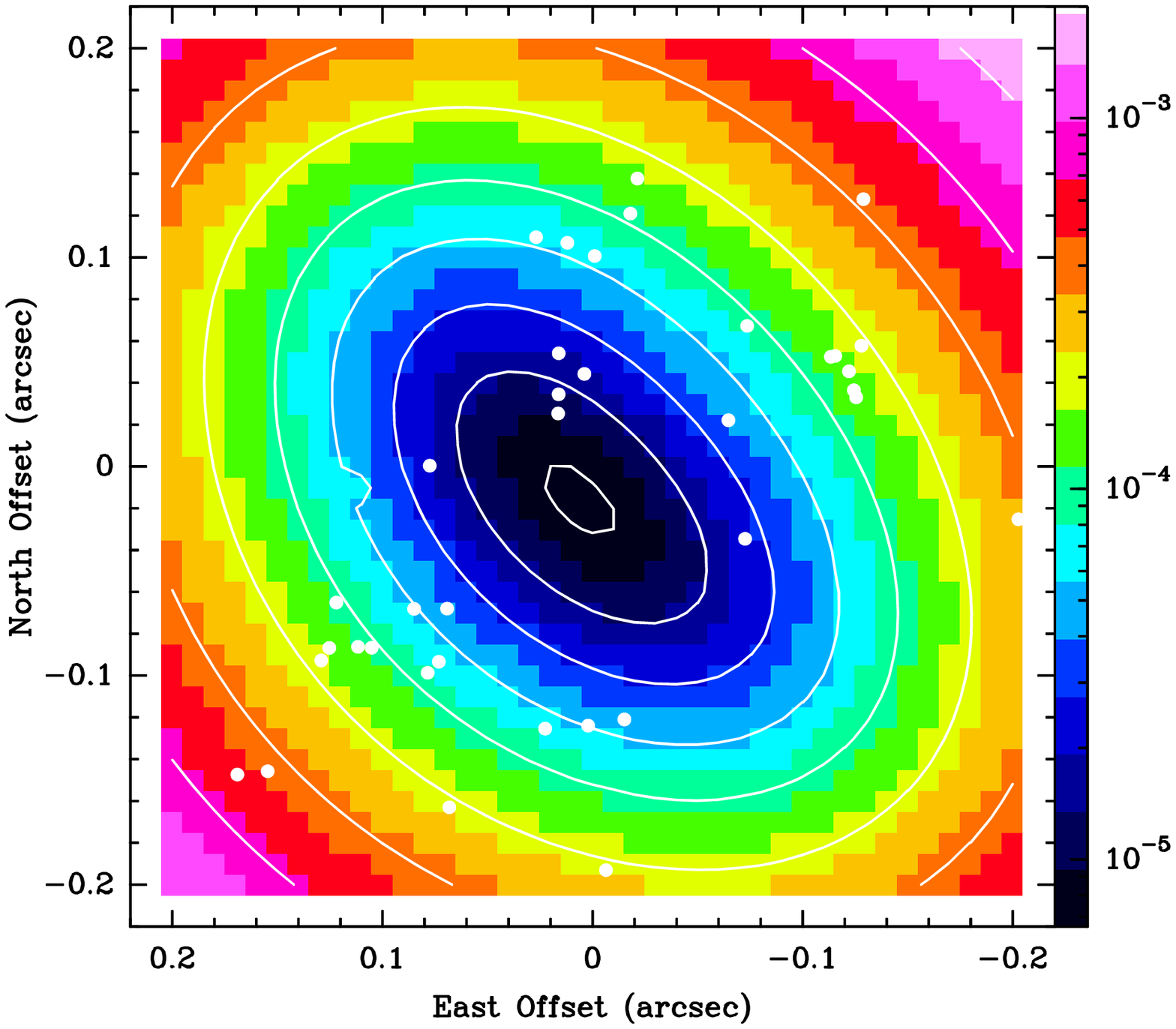}
\hspace{-2cm}
\includegraphics[width=6.3cm, angle=-90.0]{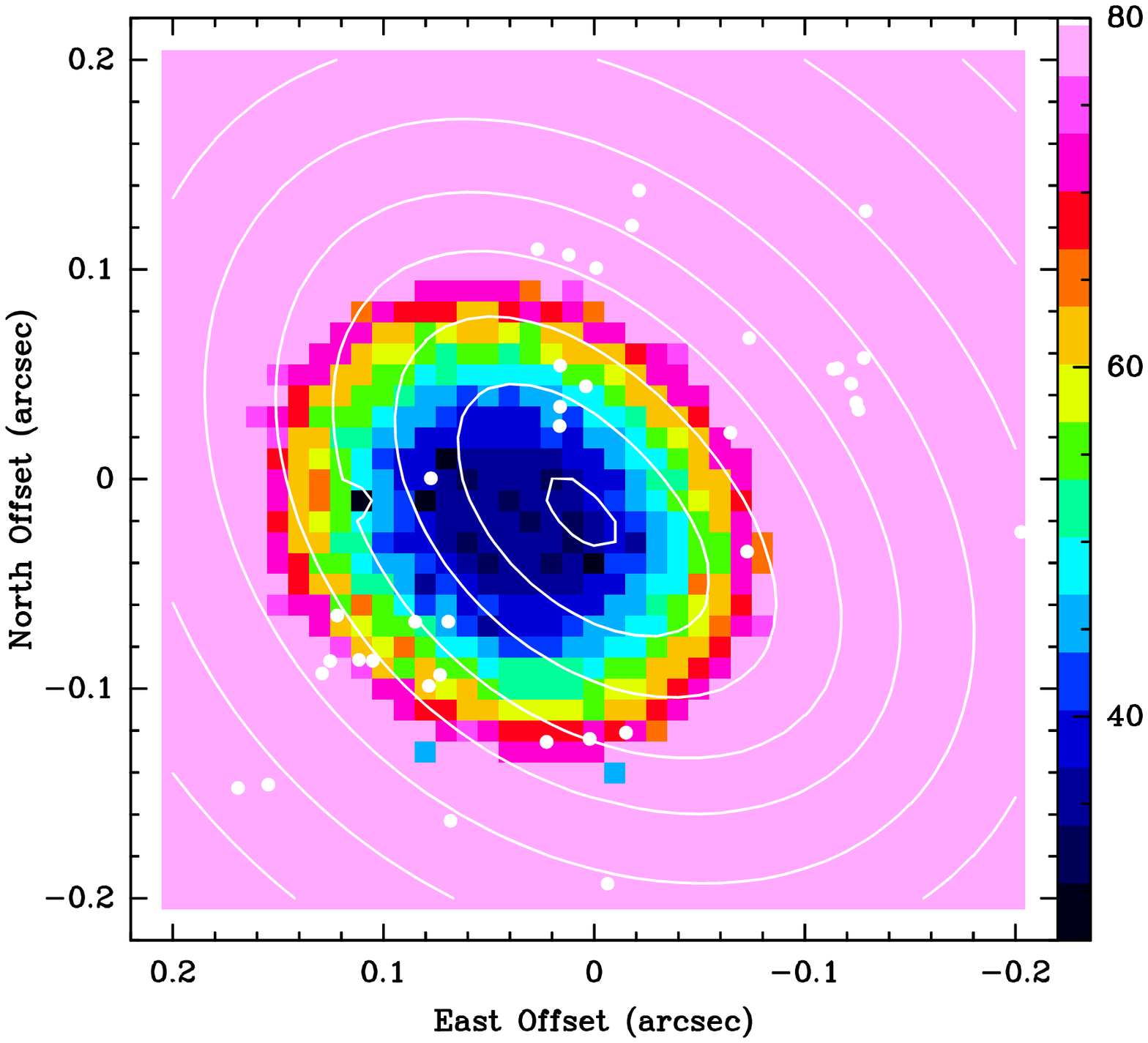}
\caption{Maps of the (mean squared) residual ({\it  left panel})
and best-fit central mass ({\it right panel})
obtained by fitting the curve \mbox{$R_{s,p}^3 = K \, / \, \Gamma^2$} to the 6.7~GHz maser data in \Gs.
{\it Colours} give the map value, with the value-color
conversion code given in the wedge on the right side of the panels. 
The values reported in the wedge are in units of \ arcsec$^6$ \ and \ solar~mass (M$_{\odot}$) \ 
for the left and right panels, respectively.
In both panels, positions
are given relative to the center of motion of the 6.7~GHz masers (as defined in SMC1).
To better compare the two maps, in both panels {\it white contours} report the same levels
of fit (mean squared) residual. The plotted levels are
from \ 7~10$^{-6}$~arcsec$^6$ (10\% higher than the minimum fit residual) to \ 1.6~10$^{-3}$~arcsec$^6$, 
stepped by a factor of 2.
The spatial distribution of the 6.7~GHz maser features is indicated with {\it white dots}.
\label{res_mas}}
\end{figure*}

Owing to the small inclination with the plane of the sky of the maser disk/toroid observed in \Gs, 
the sky-projected distances, $R_{s,p}$, should approximate the real 3-D distances, $R_{s}$, within 15\%.
Hence, we plotted the cube of the sky-projected maser distance to the (proto)star, $R_{s,p}^3$,
versus the square of the \Vlsr gradient amplitude, $\Gamma^2$, and fit to the data the curve
\ \mbox{$R_{s,p}^3 = K \, / \, \Gamma^2$}, where \ $K$ \ is a constant. In the assumption that
the maser internal gradients trace Keplerian rotation, such a fit should produce minimal residuals
when the sky-projected position of the (proto)star matches the real position. We searched for the
protostar position over a grid centered on the 6.7~GHz maser center of motion (i.e. the (proto)star location
independently estimated by averaging the position of the persistent maser features),
with a grid semi-size of 200~mas at steps of 10~mas. Figure~\ref{res_mas} 
presents the map of the fit residuals (mean squared) and best-fit for the (proto)stellar mass values.
The position of the minimal fit residual ((East, North)$=$(0,-0.02)~arcsec) is found 
at an offset of only 20~mas (to the south) from the center of motion
(at the origin of the maps shown in Fig.~\ref{res_mas}). At the position of the
minimal fit residual, the best-fit central mass (corrected for the inclination angle \ $i=30\degr$)
is \ 36~M$_{\odot}$, in optimal agreement with the value of \ 35~M$_{\odot}$ derived from the 3-D velocity field
of the 6.7~GHz masers, assuming centrifugal equilibrium (SMC1).

Thus, Fig.~\ref{res_mas} shows that the maser internal
gradients constrain the star position and mass to similar values as deduced from the maser spatial and
velocity distributions. This result strongly suggests that, in \Gs, the 6.7~GHz maser gradients have
a kinematical origin and are tracing the same (Keplerian) rotation marked by the large-scale
maser kinematics.


\subsection{Caveats}

A more quantitative comparison
between the ordered, large-scale motions traced with the
(water and methanol) masers in
each of the target sources, and
the properties (direction and amplitude) of the 6.7~GHz maser
\Vlsr gradients, is complicated by the limited knowledge
of the maser geometry,
and by the uncertainty in the position and mass of the (proto)star.
Maser emissions often sample only a few directions,
irregularly distributed across the (proto)stellar environment.
Therefore, the (proto)star position on the plane of the sky, derived
as the symmetry center  of the maser spatial and velocity distribution,
can be generally determined with an accuracy not better than \ 50--100~mas (MCR).

Furthermore, one cannot exclude that the maser \Vlsr gradients reflect 
not just a single, large-scale ordered motion (as the simple cases of either outflow or rotation discussed in Sects.~\ref{outflow}~and~\ref{disk}),
but a combination of them.
That certainly would make it harder to interpret
the maser internal gradients and to correlate them with the  kinematics of the gas close to the (proto)star(s).  
The sources \Gp\ and \IR\ can be good examples of a rather complex
kinematical pattern of the 6.7~GHz methanol masers, which trace rotation (about the protostellar jet) and outflow (across and along the jet). 
Assessment of the potential of the new tool provided by the 6.7~GHz \Vlsr gradients
to constrain the gas kinematics would certainly benefit from an increased statistics of well-studied high-mass star-forming regions.  

\section{Conclusion}

This work presents a study at high-velocity resolution of the milliarcsecond structure
of the 6.7~GHz \meth\ masers observed towards four high-mass star-forming regions:
\Gs, \Gp, \IR, and \AF. {\it Most} of the detected 6.7~GHz maser features present
an ordered (linear or arc-like) distribution of maser spots on the plane of the sky,
together with a regular variation in the spot \Vlsr with position measured along the
major axis of the feature elongation. A feature \Vlsr gradient can be defined
as a vector quantity, characterized by an orientation (that of the feature
major axis) and by an amplitude (equal to the derivative of the spot \Vlsr
with position). Typical values for the amplitude of the 6.7~GHz maser \Vlsr gradients
are found to be \ 0.1--0.2~km~s$^{-1}$~mas$^{-1}$.
Our multi-epoch VLBI observations show that, in each of the four target sources,
the orientation and the amplitude of most of the feature \Vlsr gradients
remain remarkably stable in time, on timescales of (at least) several years.
In three (\Gs, \Gp, and \IR) of the four sources under examination,
the directions of the feature gradients on the plane of the sky
concentrate in two intervals of P.A., each $\approx$30$\degr$--40$\degr$ wide,
separated by about \ 90$\degr$--100$\degr$. The two groups of features
with \Vlsr gradients oriented approximately perpendicular to each other,
present a similar spatial distribution over the whole maser region.

The observed time persistency of the 6.7~GHz maser \Vlsr gradients and the
order found in their angular and spatial distributions suggests
a kinematical interpretation for their origin. That is supported further 
by the finding that, in each of the four sources,
the data are consistent with having the \Vlsr gradients and proper motion vectors
in the same direction on the sky, considered the measurement uncertainties.
We discuss the case that the observed \Vlsr gradients can reflect
the ordered, large-scale ($\sim$100--1000~AU) motions (outflow, rotation, and infall)
traced by the (22~GHz water and 6.7~GHz methanol) masers in these regions,
on much smaller linear ($\sim$10~AU) scales.
For the source \Gs, where the 6.7~GHz masers trace a well-defined rotating structure,
we have used the maser \Vlsr gradients to constrain
the (proto)star position and mass and find good agreement with the values
derived from the (3-D) maser spatial and velocity distribution.

Our conclusion is that the study of the 6.7~GHz maser gradients
on milliarcsecond scales proves to be a useful tool
for investigating the gas kinematics in proximity to massive (proto)stars.
In the future, we plan to extend the analysis of the 6.7~GHz maser \Vlsr gradients to
other sources, with the particular aim of further investigating their origin
and, ultimately, employing them for kinematical studies.

\begin{acknowledgements}
We are grateful to Malcolm Walsmley for useful discussions
about the nature of the 6.7~GHz maser \Vlsr gradients.
This work was partially funded by the ERC Advanced Investigator Grant GLOSTAR (247078).
\end{acknowledgements}

\bibliographystyle{aa}
\bibliography{biblio}

\begin{thebibliography}{14}
\expandafter\ifx\csname natexlab\endcsname\relax\def\natexlab#1{#1}\fi

\bibitem[{{Beltr{\'a}n} {et~al.}(2011){Beltr{\'a}n}, {Cesaroni}, {Neri}, \&
  {Codella}}]{Bel11}
{Beltr{\'a}n}, M.~T., {Cesaroni}, R., {Neri}, R., \& {Codella}, C. 2011, \aap,
  525, A151+

\bibitem[{{Bloemhof} {et~al.}(1992){Bloemhof}, {Reid}, \& {Moran}}]{Blo92}
{Bloemhof}, E.~E., {Reid}, M.~J., \& {Moran}, J.~M. 1992, \apj, 397, 500

\bibitem[{{Brunthaler} {et~al.}(2009){Brunthaler}, {Reid}, {Menten}, {Zheng},
  {Moscadelli}, \& {Xu}}]{Bru09}
{Brunthaler}, A., {Reid}, M.~J., {Menten}, K.~M., {et~al.} 2009, \apj, 693, 424

\bibitem[{{Cesaroni} {et~al.}(2006){Cesaroni}, {Galli}, {Lodato}, {Walmsley},
  \& {Zhang}}]{Ces06}
{Cesaroni}, R., {Galli}, D., {Lodato}, G., {Walmsley}, M., \& {Zhang}, Q. 2006,
  \nat, 444, 703

\bibitem[{{Cragg} {et~al.}(2005){Cragg}, {Sobolev}, \& {Godfrey}}]{Cra05}
{Cragg}, D.~M., {Sobolev}, A.~M., \& {Godfrey}, P.~D. 2005, \mnras, 360, 533

\bibitem[{{Fish} {et~al.}(2006){Fish}, {Brisken}, \& {Sjouwerman}}]{Fis06a}
{Fish}, V.~L., {Brisken}, W.~F., \& {Sjouwerman}, L.~O. 2006, \apj, 647, 418

\bibitem[{{Fish} \& {Reid}(2006)}]{Fis06b}
{Fish}, V.~L. \& {Reid}, M.~J. 2006, \apjs, 164, 99

\bibitem[{{Fish} \& {Sjouwerman}(2007)}]{Fis07a}
{Fish}, V.~L. \& {Sjouwerman}, L.~O. 2007, \apj, 668, 331

\bibitem[{{Goddi} {et~al.}(2005){Goddi}, {Moscadelli}, {Alef}, {Tarchi},
  {Brand}, \& {Pani}}]{God05}
{Goddi}, C., {Moscadelli}, L., {Alef}, W., {et~al.} 2005, \aap, 432, 161

\bibitem[{{Goddi} {et~al.}(2007){Goddi}, {Moscadelli}, {Sanna}, {Cesaroni}, \&
  {Minier}}]{God07}
{Goddi}, C., {Moscadelli}, L., {Sanna}, A., {Cesaroni}, R., \& {Minier}, V.
  2007, \aap, 461, 1027 (GMS1)

\bibitem[{{Goddi} {et~al.}(2011){Goddi}, {Moscadelli}, \& {Sanna}}]{God11}
{Goddi}, C., {Moscadelli}, L., \& {Sanna}, A. 2011, submitted to A\&A (GMS2)

\bibitem[{{Matthews} {et~al.}(2010){Matthews}, {Greenhill}, {Goddi},
  {Chandler}, {Humphreys}, \& {Kunz}}]{Mat10}
{Matthews}, L.~D., {Greenhill}, L.~J., {Goddi}, C., {et~al.} 2010, \apj, 708,
  80

\bibitem[{{Moscadelli} {et~al.}(2011){Moscadelli}, {Cesaroni}, {Rioja},
  {Dodson}, \& {Reid}}]{Mos11}
{Moscadelli}, L., {Cesaroni}, R., {Rioja}, M.~J., {Dodson}, R., \& {Reid},
  M.~J. 2011, \aap, 526, A66+ (MCR)

\bibitem[{{Moscadelli} {et~al.}(2003){Moscadelli}, {Menten}, {Walmsley}, \&
  {Reid}}]{Mos03}
{Moscadelli}, L., {Menten}, K.~M., {Walmsley}, C.~M., \& {Reid}, M.~J. 2003,
  \apj, 583, 776

\bibitem[{{Sanna} {et~al.}(2010{\natexlab{a}}){Sanna}, {Moscadelli},
  {Cesaroni}, {Tarchi}, {Furuya}, \& {Goddi}}]{San10a}
{Sanna}, A., {Moscadelli}, L., {Cesaroni}, R., {et~al.} 2010{\natexlab{a}},
  \aap, 517, A71+ (SMC1)

\bibitem[{{Sanna} {et~al.}(2010{\natexlab{b}}){Sanna}, {Moscadelli},
  {Cesaroni}, {Tarchi}, {Furuya}, \& {Goddi}}]{San10b}
{Sanna}, A., {Moscadelli}, L., {Cesaroni}, R., {et~al.} 2010{\natexlab{b}},
  \aap, 517, A78+ (SMC2)

\end{thebibliography}

\clearpage

\end{document}